\documentclass[fleqn,usenatbib]{mnras}

\usepackage{newtxtext,newtxmath}

\usepackage[T1]{fontenc}

\DeclareRobustCommand{\VAN}[3]{#2}
\let\VANthebibliography\thebibliography
\def\thebibliography{\DeclareRobustCommand{\VAN}[3]{##3}\VANthebibliography}

\usepackage{graphicx}	
\usepackage{amsmath}	
\usepackage{academicons} 
\usepackage{xcolor}
\usepackage{url}
\usepackage[utf8]{inputenc}
\usepackage{hyperref}
\usepackage{orcidlink}
\usepackage{multicol}
\usepackage{pdflscape}

\newcommand{\dgr}{$^{\circ}~$}

\newcommand{\teffe}{$\rm T_{eff}$}

\newcommand{\logge}{$\log{g}$}

\title[LMC Gradients]{Revealing the Chemical Structure of the Magellanic Clouds with APOGEE. II. Abundance Gradients of the Large Magellanic Cloud}

\author[Povick et al.]{
\noindent
Joshua T. Povick\orcidlink{0000-0002-6553-7082}$^{1}$\thanks{E-mail: joshua.povick@montana.edu},
David L. Nidever\orcidlink{0000-0002-1793-3689}$^{1}$,
Steven R. Majewski\orcidlink{0000-0003-2025-3147}$^{2}$,
Doug Geisler$^{3,4,5}$,
\newauthor
 Maria-Rosa L. Cioni\orcidlink{0000-0002-6797-696X}$^{6}$,
Yuxi(Lucy) Lu\orcidlink{0000-0003-4769-3273}$^{7,8}$,
Ricardo Mu\~{n}oz$^{9}$,
Guy S. Stringfellow\orcidlink{0000-0003-1479-3059}$^{10}$,
\newauthor
 Andr\'{e}s Almeida$^{2}$,
Pen\'{e}lope Longa-Pe\~{n}a$^{11}$,
Richard R. Lane$^{12}$
and Alexandre Roman-Lopes\orcidlink{0000-0002-1379-4204}$^{5}$\\
$^{1}$Department of Physics, Montana State University, P.O. Box 173840, Bozeman, MT 59717-3840\\
$^{2}$Department of Astronomy, University of Virginia, Charlottesville, VA 22904-4325, USA \\
$^{3}$Departmento de Astronom\'{i}a, Universidad de Concepci\'{o}n, Casilla 160-C Concepci\'{o}n, Chile \\
$^{4}$Instituto de Investigaci\'{o}n Multidisciplinario en Ciencia y Tecnolog\'{i}a, Universidad de La
Serena, Avenida Ra\'{u}l Bitr\'{a}n S/N, La Serena, Chile\\
$^{5}$Departamento de Astronom\'{i}a, Facultad de Ciencias, Universidad de La Serena. Av.
Juan Cisternas 1200, La Serena, Chile \\
$^{6}$Leibniz-Institut f\"{u}r Astrophysik Potsdam, An der Sternwarte 16, D-14482 Potsdam, Germany \\
$^{7}$American Museum of Natural History, Central Park West, Manhattan, NY, USA \\
$^{8}$Department of Astronomy, Columbia University, 550 West 120th Street, New York, NY, USA \\
$^{9}$Departamento de Astronom\'{i}a, Universidad de Chile, Camino del Observatorio 1515, Las Condes, Santiago, Chile\\
$^{10}$Center for Astrophysics and Space Astronomy, University of Colorado, 389 UCB, Boulder, CO 80309-0389, USA \\
$^{11}$Centro de Astronomia, Universidad de Antofagasta, Avenida Angamos 601, Antofagasta 1270300, Chile \\
$^{12}$Centro de Investigaci\'{o}n en Astronom\'{i}a, Universidad Bernardo O'Higgins, Avenida Viel 1497, Santiago, Chile \\
}

\date{Accepted XXX. Received YYY; in original form ZZZ}

\pubyear{2023}

\begin{document}
\label{firstpage}
\pagerange{\pageref{firstpage}--\pageref{lastpage}}
\maketitle

\begin{abstract}

    We present the abundance gradients of the Large Magellanic Cloud (LMC) for 25 elemental abundance ratios and their respective temporal evolution as well as age-[X/Fe] trends using 6130 LMC field red giant branch (RGB) stars observed by SDSS-IV / APOGEE-2S. APOGEE is a high resolution ($R$ $\sim$22,500) $H$-band spectroscopic survey that gathered data on the LMC with broad radial and azimuthal coverage out to $\sim$10\degr. The calculated overall metallicity gradient of the LMC with no age binning is $-$0.0380 $\pm$ 0.0022 dex/kpc. We also find that many of the abundance gradients show a U-shaped trend as functions of age. This trend is marked by a flattening of the gradient but then a general steepening at more recent times. The extreme point at which all these gradients (with the U-shaped trend) begin to steepen is $\gtrsim$2 Gyr ago. In addition, some of the age-[X/Fe] trends show an increase starting a few Gyr before the extreme point in the gradient evolutions. A subset of the age-[X/Fe] trends also show maxima concurrent with the gradients' extreme points, further pinpointing a major event in the history of the LMC $\sim$2 Gyr ago. This time frame is consistent with a previously proposed interaction between the Magellanic Clouds suggesting that this is most likely the cause of the distinct trend in the gradients and age-[X/Fe] trends.

\end{abstract}




\section{Introduction}
\label{sec:lmc_intro}

While much progress has been made in our understanding of galaxy formation and evolution over the past several decades, both from the observational and theoretical sides, there are still important aspects that remain to be solved.
It is widely accepted that the first galaxies formed from material falling into dark matter halos creating smaller ``dwarf'' galaxies that later merge with other dwarfs to create larger galaxies such as the Milky Way (MW)
(\citealt{white1978merge}; \citealt{searle78}). 
Given this hierarchical galaxy formation,
a key step 
to understanding how larger galaxies form is 
examining the properties and evolution of dwarf galaxies.
In particular, their elemental abundances and distributions 
hold powerful clues to the chemical evolution of their future host systems.


While dwarf galaxies are the most abundant galaxies in the universe, these systems are 
intrinsically faint and even most of the closest 
are 
far away, making them difficult to study.
Fortunately, the Local Group is rife
with dwarf galaxies, which brings a large number that are close enough to resolve individual stars and study in great detail \citep[e.g.,][]{mateo1998localgroup,mcconnachie2012localgroup}. The largest satellites of the MW
are the Large and Small Magellanic Clouds (LMC and SMC, respectively).
Both of the Magellanic Clouds (MCs) are relatively close, with the LMC
at 49.9 kpc \citep{degrijs2014clustering,van2001magellanic,pietrzynski2019lmcdistance} and the SMC at 62.44 kpc \citep{graczyk2020distance}. 
The MCs have been studied for decades with wide-field optical \citep[e.g.,][]{nidever2017smash}, NIR \citep[e.g.,][]{Skrutskie2006} and radio surveys \citep[e.g.,][]{Staveley-Smith2003}.  
However, wide-field, high-resolution spectroscopic surveys have only started more recently \citep[e.g.,][]{olsen2011,nidever20lazy,cullinane2020}.

The study of the LMC abundance gradients started in the 1970s with such works as \cite{Dufor1975hii} and \cite{pagel1978hii}, which looked at LMC H{\scriptsize II} regions. 
Although neither of these studies found a 
discernible gradient, the existence of a
 non-zero gradient could also not be ruled out. Since then, there have been many other studies of abundance gradients in the LMC: some of them detect a gradient \citep[e.g.][]{kontizas1993,cioni2009grad,feast2010lmcgrad,wagner2013grad}, while others do not \citep[e.g.][]{pena1987grad,geisler2003grad,grocholski2006ii,sharma2010grad,palma2015grad}.
The debate is still ongoing and one of the goals of this work is to show that the LMC does in fact have a measurable metallicity and other various abundances 
gradients.

The present study breaks new ground 
by measuring
radial gradients and their age-dependence for 25 chemical elements through use of 
high-resolution spectroscopy of thousands of LMC field stars. While it is possible to use star clusters to accomplish something similar in galaxies such as the 
MW \citep{donor2020occam}, this is not possible in the LMC. An age gap exists in LMC star clusters between $\sim$3$-$12 Gyr \citep{dacosta1991clusters,geisler1997oldclusters} with very few clusters having been identified in the gap \citep{mateo1986ccd,olszewski1991spectroscopy,piatti2022genuine}. Clearly this does not bode well for exploring the full evolution of gradients derived from clusters with so much ``missing'' temporal information. Fortunately, the LMC contains many field stars for which ages can be derived and this is the approach we take for our study.

In Povick et al. (2023a --- in preparation; hereafter Paper I), we described our method for calculating ages for individual LMC field stars.  Spectroscopic parameters (\teffe,
\logge, [Fe/H], and [$\alpha$/Fe]), multi-band photometry, and isochrones are used to accomplish this.  
Ages for individual RGB stars are found by 
calculating apparent isochrone magnitudes for a star using the spectroscopic parameters (\teffe, \logge, [Fe/H], and [$\alpha$/Fe]), PARSEC isochrones \citep{Girardi2002}, and an inclined plane disk model \citep{choi2018reddening} for the stellar distances.  The age and extinction are varied until the best match with the observed photometry is found.



In this paper, we start by discussing the APOGEE data in Section \ref{sec:lmc_data}. Then, Section \ref{sec:lmc_gradmethod} outlines the inclined disk geometry and radial abundance trend calculation. Section \ref{sec:lmc_results} presents the results. From there we discuss the implications of the results in Section \ref{sec:lmc_discussion} and, finally, our main conclusions are summarized in Section \ref{sec:lmc_summary}.


\section{APOGEE Data}
\label{sec:lmc_data}

The spectroscopic data for this work comes from the Apache Point Observatory Galactic Evolution Experiment \citep[APOGEE,][]{majewski2017apache}. More precisely, the data were taken by the second phase
of the survey (APOGEE-2) which, for the first time, obtained data in the Southern Hemisphere (APOGEE-2S) as
part of the Sloan Digital Sky Survey-IV (SDSS-IV; \citealt{blanton2017sloan}). All of the data are available as part of the public SDSS-IV Data Release 17 \citep[DR17,][]{Abdurro'uf2022ApJSdr17}.

The APOGEE survey was designed to study the chemical enrichment and kinematics of the MW
with wide Galactic coverage. The survey is dual hemisphere with two identical \textit{H}-band spectrographs \citep{wilson19spectro} attached to the Sloan 2.5-m Telescope at the Apache Point Observatory \citep[APO,][]{gunn2006apo} 
in New Mexico and to the 2.5-m du Pont telescope at the Las Campanas Observatory \citep[LCO,][]{bowen1973optical} in Chile. 
Information regarding the targeting of the telescopes for the survey can be found in \cite{zasowski13target} and \cite{zasowski17target}.

After observations were 
obtained, the stellar spectra were
initially reduced with the APOGEE reduction pipeline \citep{nidever15pipe}. After this first reduction, the stellar parameters were derived using the APOGEE Stellar Parameter and Chemical Abundance Pipeline \citep[ASPCAP,][]{holtzman15abund,perez16aspcap}. ASPCAP works by using the \texttt{FERRE} \citep{allendeprieto2006} software to compare the observed spectra to a library of synthetic spectra created using \texttt{synspec} \citep{hubeny21synspec}.  The library is searched for the best-matching synthetic spectrum and, thereby, the main stellar parameters that affect the global spectrum (\teffe, $\log{g}$, v$_{\text{micro}}$, [M/H], [C/M], [N/M], [$\alpha$/M], and v$_{\text{macro}}$) are determined. Finally, ASPCAP derives abundances for C, C{\scriptsize I}, N,
O, Na, Mg, Al, Si, S, K, Ca, Ti, Ti{\scriptsize II}, V, Cr, Mn, Fe, Co, Ni,
and Ce by holding the stellar parameters fixed and only varying the [M/H] dimension (or [$\alpha$/M] dimension for the $\alpha$-elements) and finding the best-fitting synthetic spectrum using wavelength windows unique to each element. The lines lists used by ASPCAP can be found in \cite{smith2021lines} and \cite{shetrone2015lines}.  Most of these abundances and various abundance ratios are studied in this work. For more information on the data reduction process and updates see \cite{jonsson2020data} and specifically for DR17 see Holtzman et al.\ (in preparation). It should also be noted that Na, K, P, S, V, Cu, and Ce can suffer from higher noise than the rest of elements and may have less reliable measurements in the catalog.

Like all observations, APOGEE measurements suffer from statistical uncertainties as well as systematic bias, which must be accounted for. To mitigate the problems that arise due to systematic biases, APOGEE uses solar neighborhood stars with known solar-like metallicities and derived offsets that were applied to the APOGEE parameters and abundances. Statistical uncertainties were determined using multiple visits for stars and the scatter was fit as a function of \teffe, [M/H], and SNR.\footnote{The goal for each APOGEE spectrum is to reach an SNR of 100 per half-resolution unit, though stars with SNR $<$ 100 are not necessarily discarded \citep{jonsson2020data}.} The uncertainties were then somewhat inflated to be more in line with those seen for first generation stars in globular clusters. More details on the uncertainties specific to
the Magellanic Cloud stars can be found in \cite{nidever20lazy}.

\begin{figure}
    \centering
    \includegraphics[scale=0.31]{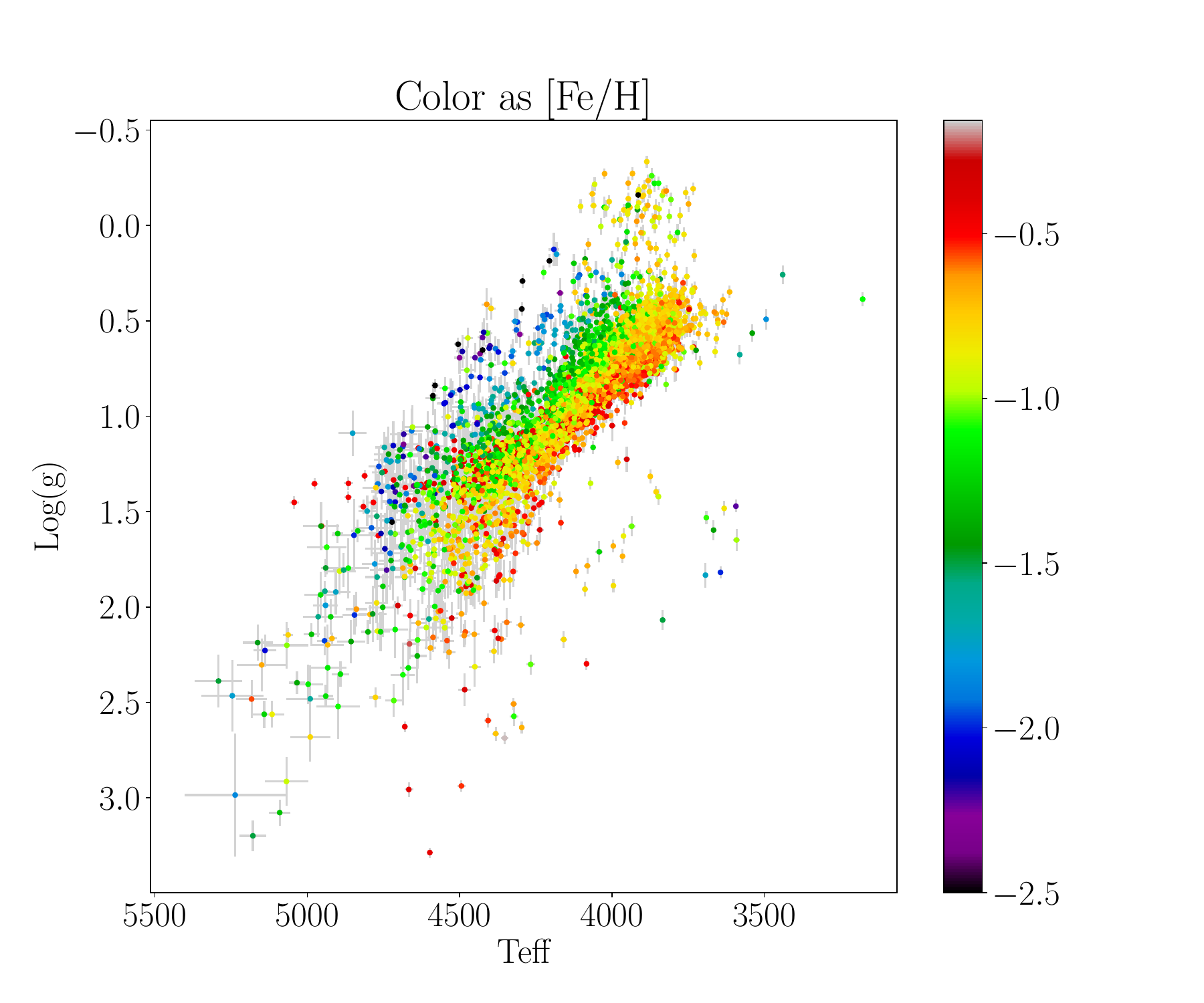}
    \caption{A Kiel-diagram for the whole APOGEE DR17 sample of the LMC RGB stars with associated uncertainties. The color of each of the stars in the plot represent their respective APOGEE metallicities.}
    \label{fig:lmc_HR}
\end{figure}

In the end, APOGEE was able to observe and measure abundances for 6130 red giant branch (RGB) stars in 
36 LMC fields. A Kiel-diagram of the full APOGEE-2S LMC RGB sample is presented in Figure \ref{fig:lmc_HR}. APOGEE has wide 
azimuthal and radial coverage out to $\sim$10\dgr, as can be seen in Figure \ref{fig:lmc_lmcgeo}.

\subsection{BACCHUS Supplemental Data}
\label{ssec:lmc_bacchus}

In \cite{hayes2022bacchus}, the APOGEE DR17 spectra of 120,000 giants were reanalysed to derive abundances for elements that possess weak or blended spectral lines in the APOGEE data. Using the Brussels Automatic Code for Characterizing High-accuracy Spectra (\texttt{BACCHUS}) abundances were derived for Na, P, S, V, Cu, Ce, and Nd as well as the $^{12}$C/$^{13}$C isotopic ratio.  The \texttt{BACCHUS} abundances for these weak-lined elements are generally more accurate than those produced by ASPCAP. In addition, the ASPCAP P, Cu, and Nd abundances are not reliable nor contained in the DR17 data release, so these are provided only through the \texttt{BACCHUS} study of the DR17 APOGEE spectra.

The \texttt{BACCHUS} catalog includes data for 763 of the LMC RGB stars.
The present study makes use of the Ce and Nd abundances derived with the \texttt{BACCHUS} code. While the main APOGEE catalog does derive Ce abundances for many more stars \citep{cunha2017adding}, the evolutionary trend of the Ce gradient is slightly more coherent when using the \cite{hayes2022bacchus} abundances versus the ASPCAP ones and the \texttt{BACCHUS} derived Ce values typically a $\sim$0.03 dex reduction in the error values. 
With the exception of Nd, none of the other \texttt{BACCHUS} abundances have enough stars with reliable values to calculate good gradients.

\section{Abundance Gradient Calculation}
\label{sec:lmc_gradmethod}

\subsection{LMC Geometry}
\label{ssec:lmc_geometry} 

\begin{figure}
    \centering
    \includegraphics[width=0.45\textwidth]{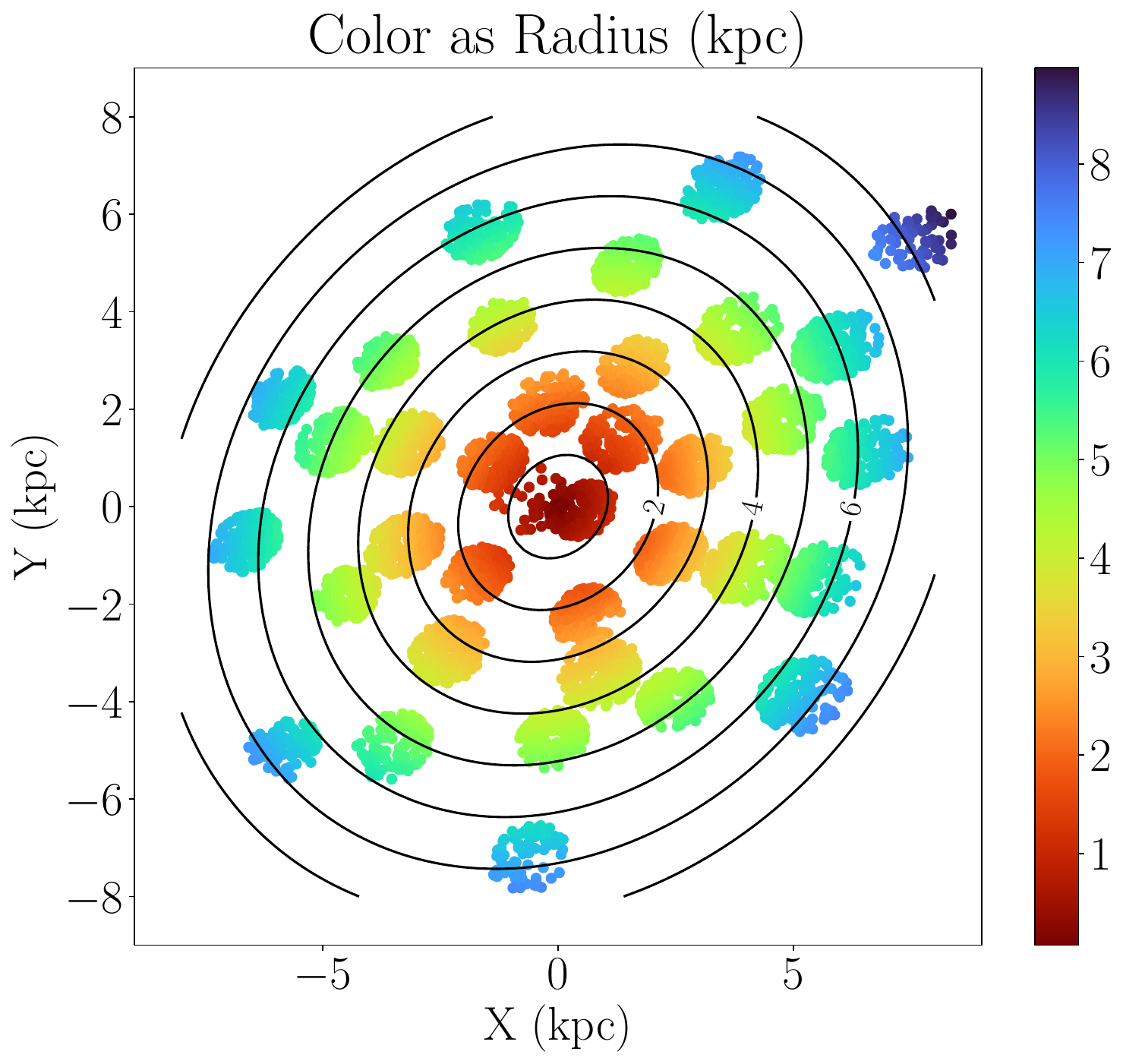}
 \caption{The resulting distance map for the LMC using Equation \ref{equ:ellrad} with the color indicating radial distance from the center projected onto a tangential Cartesian plane in kpc. The contours show lines of constant ``elliptical'' radius starting at 1 kpc for the center contour out to 8 kpc calculated using Equation 6 from \protect\cite{choi2018smashing}. For reference the center of the map, (X, Y) = (0, 0), is at ($\alpha$, $\delta$) = (05:27:36,$-$69:52:00).}
    \label{fig:lmc_lmcgeo}
\end{figure}

The geometry of the LMC disk is well-modeled by an inclined elliptical disk.  The mathematical framework was developed by \cite{van2001magellanic} and \cite{choi2018smashing} and consists of two steps to convert equatorial coordinates to a Cartesian coordinate system with an origin at the center of the LMC. First the equatorial coordinates ($\alpha$, $\delta$) are shifted so that the origin is at the center of the LMC and second this coordinate system is then projected onto a tangent plane. Effectively this creates a 2D coordinate system for the disk of the galaxy. It is also possible to get distance from this transformation creating a third dimension, though the distance is not used herein except for deriving ages (see Paper I). Since the LMC is elliptical, this must be considered when determining the radius. We take the equation for finding the radius of an ellipse directly from \cite{choi2018smashing}, which is
\begin{equation}
\label{equ:ellrad}
    r(x,y)^2 = (x\cos{\psi} - y\sin{\psi})^2 + \bigg(\frac{x\sin{\psi} + y\cos{\psi}}{b/a}\bigg)^2,
\end{equation}
\noindent
where $\psi\,(= 227.24^\circ)$ is the position angle of the semi-major axis, and $b/a\,(= 0.836)$ is the ratio of the semi-minor axis to the semi-major axis.  Graphically, the geometry of the LMC can be seen in Figure \ref{fig:lmc_lmcgeo}. Also for reference, the disk scale length for this model is 1.667 $\pm$ 0.002 kpc \citep{choi2018smashing}.

\subsection{MCMC}
\label{ssec:lmc_mcmc}

Radial abundance trends can be found assuming a simple linear model that is only functionally dependent on the radius from the center of the LMC. Using a linear model is the convention for determining abundance gradients, but in reality a line is not necessarily the best model; thus a linear model will give the abundance gradient only to a first order approximation. The model used to find the abundance gradients is given by

\begin{equation}
\label{equ:linmodl}
    [\text{X/Fe}] = \nabla_{R}R+[\text{X/Fe}]_0,
\end{equation}

\noindent
where $\nabla_R$ is the gradient in dex per kpc (or dex/kpc), $R$ is the projected distance from the center of the LMC, and $[\text{X/Fe}]_0$ is the central abundance. 

Most of the abundances measured by APOGEE use iron as the fiducial element as in the above Equation \ref{equ:linmodl}, but this is only for demonstration purposes.  
Gradients are also calculated using hydrogen and magnesium as fiducials. Hydrogen is an obvious choice for a fiducial as it removes removes any ``convolution'' with any other metal and gives an absolute abundance gradient. In a sense [X/H] more like ``first order'' chemical evolution effects, while any other fiducial offer a glimpse at ``second order'' chemical evolution effects. Magnesium was also chosen as a fiducial because it almost entirely enriches the interstellar medium (ISM) through core collapse supernovae with only the slightest [Fe/H] dependence \citep{andrews2017inflow}. On the other hand, iron enriches the ISM through multiple channels and, therefore, its relationship to other elements is more complicated compared to magnesium \citep{weinberg2019chemical}. 
[C/N] and [$\alpha_\text{h}$/$\alpha_\text{ex}$] are also investigated and obviously these  do not have an Fe, H, or Mg fiducial. These are the only two abundances ratios for which this is true. For future reference [$\alpha_\text{h}$/$\alpha_\text{ex}$] is the iron free $\alpha$ abundance where [$\alpha_\text{h}$/Fe] is the average hydrostatic $\alpha$ abundance and [$\alpha_\text{ex}$/Fe] is the average explosive $\alpha$ abundance. Further discussion of  this can be found in Section \ref{sec:lmc_results}.



To find the parameters of the linear model in Equation \ref{equ:linmodl}, we use maximum likelihood estimation (MLE). MLE has advantages over other methods, such as ordinary least squares (OLS), because it does not assume that the uncertainties are necessarily Gaussian like OLS does. In the limit of Gaussian errors, MLE approaches OLS. 

Implementing MLE requires maximizing the likelihood function, which is the same as minimizing the negative log likelihood function. The log likelihood function for the linear model is
\begin{equation} 
\label{equ:lnL}
    \ln{\mathcal{L}} = \frac{1}{2}\bigg[ \sum_{i} \frac{([\text{X/Fe}]_i - \nabla_RR_i - [\text{X/Fe}]_0)^2}{\sigma_i^2} - \ln{2\pi \sigma_i^2} \bigg],
\end{equation}

\noindent
where $R_i$ is the projected distance from the center of the LMC to the individual star,  $[\text{X/Fe}]_i$ is its abundance, $[\text{X/Fe}]_0$ corresponds to the central abundance value (see Equation \ref{equ:linmodl}), and $\sigma_i$ is the uncertainty in the abundance of the same star.

To obtain the uncertainties for each of the model parameters, the \texttt{emcee} \citep{foreman2013emcee} Python package is used. This algorithm
implements the affine invariant Markov chain Monte Carlo ensemble sampler outlined by \citet{goodman2010ensemble}.  Thirty-two walkers are used with 5000 steps each. The first int(max($\tau$)) $+$ 50 steps are removed as ``burn-in'', where the walkers are still finding the optimal value. Here max($\tau$) is the largest of the calculated auto-correlation times for a set of parameters ($\nabla_R$, [Fe/H]$_0$). Priors are assumed to flat for all parameters that are fit. Also, the chain is thinned by selecting every 15th value.


When fitting radial trends, the individual stars are not used. Instead the stars are radially binned and the gradients are fit to the median values and the median absolute deviation (MAD) of the median abundance value is used for the uncertainty in each bin. This is done to create a more robust fit, speed up the process, and not have the gradients weighted too highly in radial regions with many stars.

For the actual fit, stars with less reliable abundances are removed by requiring S/N $>$ 100 for their spectra. Here S/N is measured per half-resolution unit (for more information see \cite{jonsson2020data}). This reduces the sample size to 3824 stars. Then the stars are binned into 1 kpc radial bins and the abundance trend is fit with \texttt{emcee}. After this, outlier stars more than 3 $\times$ MAD from the best-fit model are removed and the trend is refit to obtain the final abundance gradient. 

\subsection{Stellar Age Binning}
\label{ssec:lmc_age}

All ages were calculated using Paper I. The age of a stars is derived by using the measured APOGEE spectroscopic parameters (\teffe, \logge, [Fe/H], and [$\alpha$/Fe]) to derive model photometry (\textit{BP}, \textit{G}, \textit{RP}, \textit{J}, \textit{H}, \textit{Ks}), the previously mentioned distances, and a trial age with the aid of PARSEC isochrones \citep{bressan2012parsec,marigo2017}. The trial age is then varied until the calculated and observed photometry produce the best match.
Some stars are given ages that are clearly older than the age of the universe. Most likely this is due to noise in the stellar parameters as the age distribution appears to reach a floor at the oldest ages. It is also possible that this is compounded by some source of systematic effects due to the high degeneracy in color-magnitude of very old isochrones.

Each star was placed into one of five different age bins. The bins were chosen such that the total number of stars in each bin was roughly the same. The particular age ranges of each of these bins can be seen in Table \ref{tab:lmc_bin}. This binning scheme has good age resolution while also mitigating the effect of stars whose calculated ages are quite large. In our analysis no significant differences in abundance gradients were found when varying the age bins, unless very large deviations were attempted. Also note that the age bins were applied before the SNR > 100 cut mentioned in Section \ref{ssec:lmc_mcmc}, but applying a cut in SNR before binning in age does not cause significant differences.

\begin{table}
	\centering
	\caption{Values of the different age bins used throughout this work. N is the total number of stars in each bin and N$_{\text{SNR}>100}$ is the number of stars in each bin that have an SNR over 100. For a few stars ages could not be derived and these were cut out when binning and this leads to a discrepancy when totaling the N column and comparing to the previously stated sample size.}
	\label{tab:lmc_bin}
	\begin{tabular}{cccc} 
		\hline
		Bin & Age Range & N & N$_{\text{SNR}>100}$ \\
                & (Gyr) & \\
		\hline
		1 & t $\leq$ 2.23 & 1204 & 799\\
		2 & 2.23 $<$ t $\leq$ 3.66 & 1202 & 831 \\
		3 & 3.66 $<$ t $\leq$ 5.58 & 1207 & 819 \\
            4 & 5.58 $<$ t $\leq$ 8.36 & 1200 & 698 \\
            5 & 8.36 $<$ t  & 1202 & 582 \\
		\hline
	\end{tabular}
\end{table}


\subsection{Age-[X/Fe] Trends}
\label{ssec:lmc_age_xfe_trend}


In addition to the radial abundance ratio trends, we also explore the age-[X/Fe] trends as well as the age-[C/N] and age-[$\alpha_\text{h}$/$\alpha_\text{ex}$]. When calculating these trends a different binning scheme is used from that described in Sections \ref{ssec:lmc_mcmc} and \ref{ssec:lmc_age}. Stars were placed into three different radial bins, or annuli, with the same number of stars in each bin. This is done because the outskirts of the LMC in the APOGEE sample have significantly less stars than  the central regions. Three bins were chosen so that the inner bin contains the central disk, the outer bin contains the ``edge'' of the disk, and the middle bin probes intermediate radii. The radial bins selected were: R $<$ 3.3 kpc, 3.3 $\leq$ R $<$ 4.86 kpc, and 4.8 $<$ R kpc. The annuli were further subdivided into north and south based on the position angle of the stars. A north-south division was chosen because the  LMC bar roughly run across the LMC splitting the galaxy in half. There are also known differences between the radial velocities of stars \citep{feitzinger1979velocity} and optical depth \citep{Subramanian2009depth} when comparing the northern and southern parts of the galaxy. This motivates exploring if the chemistry also differs between the two halves of the galaxy.






For each of the positional bins, age-[X/Fe] trends were calculated by splitting the stars into age bins and determining the median abundance and MAD scatter. 
The number of age bins for the age-[X/Fe] trends is calculated using $N_\text{bins} = \bigg\lceil \sqrt{N_\text{data points}}\bigg\rceil$. 
The number of age bins varies somewhat with element and were chosen to provide fairly smooth trends but without sacrificing too much temporal resolution.

\begin{figure}
    \centering
    \includegraphics[width=0.45\textwidth]{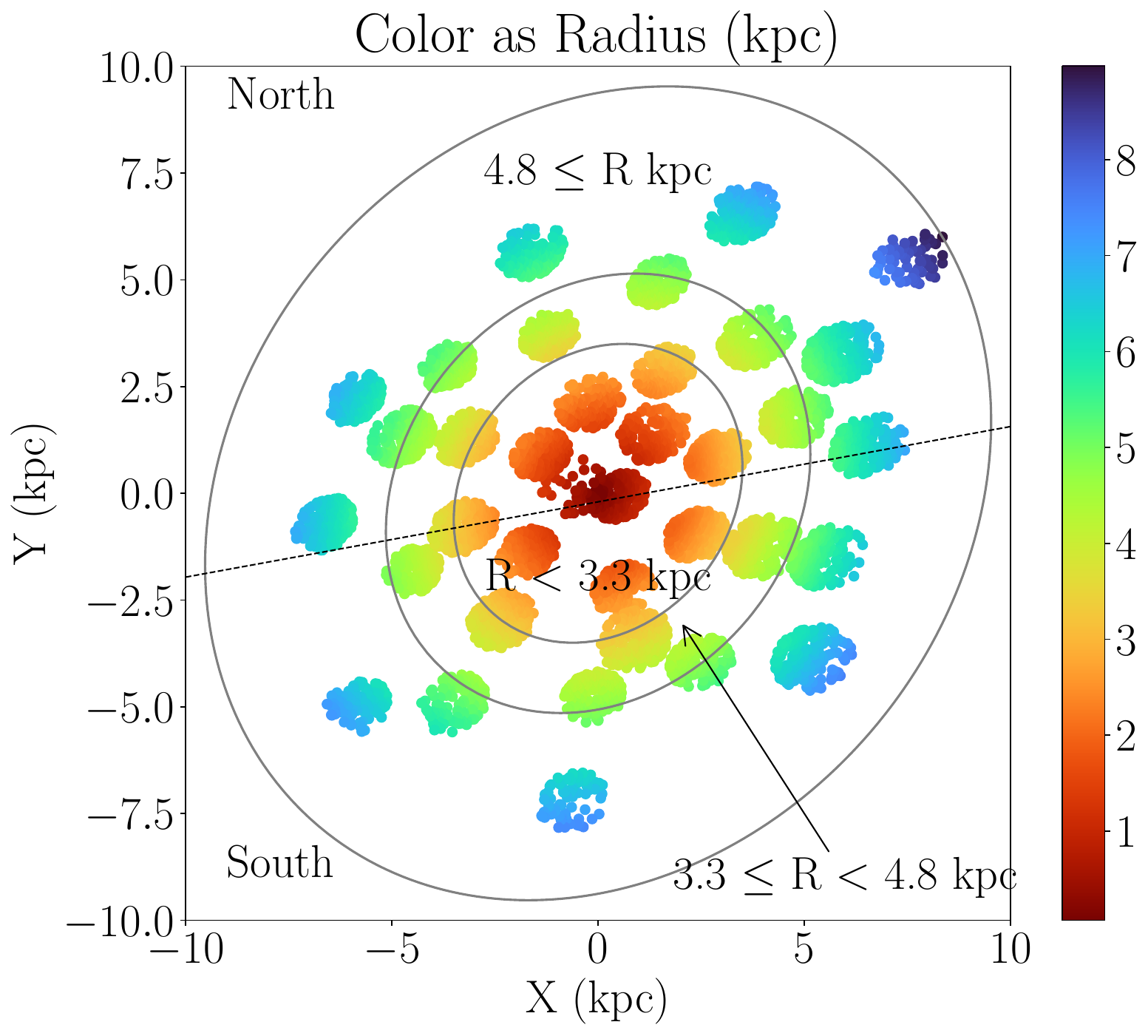}
    \caption{A plot showing the radial bins used to calculate the age-[X/Fe] relations. The dashed black line shows the boundary between the North and South LMC using the position angle of a star. This line amounts to a delineation in declination at $\sim$ $-$70$^\circ$. The solid black lines show the boundary of the different annular bins. Each of the annuli contain the same number of stars.}
    \label{fig:lmc_annuli_axfe_relations}
\end{figure}

\section{Results}
\label{sec:lmc_results}

In this section we discuss the main results of the analysis carried out using the methods in Section \ref{sec:lmc_gradmethod}. These include the [X/Fe], [X/H], and [X/Mg] gradients and their evolution as well as the age-[X/Fe] trends.

\subsection{The LMC Radial Abundance Ratio Gradients}
\label{ssec:lmc_over_grad}

The radial abundance ratio gradients in this section represent the overall time-averaged values because no age-binning was done when they were calculated. After this section, these gradients will be referred to as the overall gradients.

\subsubsection{Carbon \& Nitrogen}
\label{ssec:lmc_cn_over}

Both carbon and nitrogen are very abundant elements whose abundances are measured by APOGEE. These elements are mostly produced in intermediate-mass stars \citep{ventura2013agb} and violent Type II SNe \citep{kobayashi2006chemical}. In addition to these elements, the C+N abundance and [C/N] are also analyzed. C and N are created through nuclear burning and are brought up from the core of a star to the surface by dredge-up, which alters 
the observed C and N abundances through vigorous convection. However, \cite{gratton2000mixing} showed that the overall C+N abundance stays relatively constant in the stellar atmospheres regardless. Because of this the addition of C+N in our study is a logical choice. In addition, [C/N] is known to be age sensitive \citep[at least in relatively metal-rich stars;][]{ness2016spectroscopic} and so it makes sense to include this abundance ratio herein, especially given the incorporation of ages from Paper I in our analysis. 


The C, N, and C+N gradients all show consistent relative behavior irrespective of whether the fiducial element is Fe, H, or Mg (see Table \ref{tab:lmc_cn_over_grad}). The nitrogen gradients tend to be very flat or at least much flatter than what is seen for C and C+N, though the [N/H] gradient is decidedly not flat with a value of $-$0.0540 $\pm$ 0.0027 dex/kpc. The similarities seen in the C and C+N gradients are most likely due to the carbon abundances dominating the weighted sum when calculating the C+N values. As for the [C/N] gradient, it is very similar to the [(C+N)/Mg] gradient with a value of $-$0.0115 $\pm$ 0.0015 dex/kpc. 
A comparison of the overall radial [X/Fe] trends for carbon and nitrogen and [C/N] is given in Section \ref{ssec:lmc_cn_evolve} where we look at the evolution of these trends. 

 


\begin{table}
	\centering
	\caption{A table of overall gradients for the abundant metals carbon and nitrogen using Fe, H, and Mg as fiducials. The [C/N] gradient is also included at the bottom.}
	\label{tab:lmc_cn_over_grad}
	\begin{tabular}{cccc}
		\hline
		Element & $\nabla_R$ [X/Fe] & $\nabla_R$ [X/H] & $\nabla_R$ [X/Mg]\\
                    & (dex/kpc) & (dex/kpc) & (dex/kpc)  \\
            \hline
            C & $-$0.0281 $\pm$ 0.0010 & $-$0.0676 $\pm$ 0.0031 & $-$0.0212 $\pm$ 0.0012 \\
            N & $-$0.0028 $\pm$ 0.0010 & $-$0.0540 $\pm$ 0.0027 &  $-$0.0002 $\pm$ 0.0009 \\
            C+N & $-$0.0232 $\pm$ 0.0008 & $-$0.0631 $\pm$ 0.0028 & $-$0.0129 $\pm$ 0.0008 \\
            \hline
            {[C/N]} & $-$0.0115 $\pm$ 0.0015 & ... & ... \\
            \hline
	\end{tabular}
\end{table}

\subsubsection{\texorpdfstring{$\alpha$}\,-Elements}
\label{ssec:lmc_alpha_over}

The $\alpha$-elements enrich the ISM primarily through Type II SNe \citep{nomoto2013nucleosynthesis,weinberg2019chemical}. The $\alpha$-abundances measured by APOGEE include O, Mg, Si, S, Ca, Ti, and mean $\alpha$. In addition to the elemental $\alpha$-abundances, we also inspected the ASPCAP [$\alpha$/Fe], and the average hydrostatic $\alpha$-abundance ratio ([$\alpha_h$/Fe]=([\text{O/Fe}]+[\text{Mg/Fe}])/2) and average explosive $\alpha$-abundance ratio ([$\alpha_\text{ex}$/Fe] = ([\text{Si/Fe}]+[\text{Ca/Fe}]+[\text{Ti/Fe}])/3) as well as the ratio of hydrostatic and explosive $\alpha$-abundances ([$\alpha_\text{h}/\alpha_\text{ex}$] = [$\alpha_h/\text{Fe}$]-[$\alpha_\text{ex}/\text{Fe}$]) or ``HEx ratio''. The hydrostatic $\alpha$-elements are produced by nuclear fusion inside the star, while the explosive $\alpha$-elements are created through the Type II SNe explosion itself \citep{carlin2018chemical}. The [$\alpha_\text{h}/\alpha_\text{ex}$] is an iron free abundance that tracks the two different modes of evolution for $\alpha$-element nucleosynthesis.  \cite{carlin2018chemical} performed a spectroscopic study of Sagittarius stream stars and showed that the HEx ratio of Sagittarius stars are deficient compared to Milky Way stars but are consistent with a top-light IMF.


The [X/Fe] gradients (Table \ref{tab:lmc_alpha_over_grad}) tend to be positive and shallow. Of these, the [S/Fe], [Ti/Fe], and [$\alpha_\text{ex}$/Fe] stand out. [S/Fe] has a relatively steep positive gradient of $+$0.0236 $\pm$ 0.0028 dex/kpc and [Ti/Fe] has a steep negative gradient of $-$0.0186 $\pm$ 0.0015 dex/kpc. [$\alpha_\text{ex}$/Fe] has a slightly negative gradient of $-$0.0036 $\pm$ 0.0006 dex/kpc, but this is most likely because it is the mean of the explosive $\alpha$-elements including Ti, which, as just mentioned, has a large negative gradient.
As for the hydrogen $\alpha$-gradients there is more homogeneity with each gradient being negative and steeper than the corresponding [X/Fe] gradients, but [Ti/H] and [$\alpha_\text{ex}$/H] stand out just like for the [X/Fe] abundance ratios. The [X/Mg] gradients for the $\alpha$-elements are similar to the [X/Fe]-gradients, though clearly not the same. The [$\alpha_\text{h}$/$\alpha_\text{ex}$] has a value of $+$0.0047 $\pm$ 0.0005 dex/kpc. 


\begin{table}
	\centering
	\caption{A table of overall gradients for the $\alpha$-elements. The [$\alpha_\text{h}$/$\alpha_\text{ex}$] gradient is included at the bottom.}
	\label{tab:lmc_alpha_over_grad}
	\begin{tabular}{cccc}
		\hline
		Element & $\nabla_R$ [X/Fe] & $\nabla_R$ [X/H] & $\nabla_R$ [X/Mg]\\
                & (dex/kpc) & (dex/kpc) & (dex/kpc)  \\
    		\hline
                O & 0.0023 $\pm$ 0.0005 & $-$0.0419 $\pm$ 0.0023 & 0.0023 $\pm$ 0.0004 \\
                Mg & 0.0021 $\pm$ 0.0007 & $-$0.0475 $\pm$ 0.0024 & ... \\
                Si & 0.0054 $\pm$ 0.0005 & $-$0.0351 $\pm$ 0.0022 & 0.0039 $\pm$ 0.0007 \\
                S & 0.0263 $\pm$ 0.0028 & $-$0.0281 $\pm$ 0.0027 & 0.0283 $\pm$ 0.0028 \\
                Ca & 0.0030 $\pm$ 0.0004 & $-$0.0383 $\pm$ 0.0022 & 0.0021 $\pm$ 0.0006 \\
                Ti & $-$0.0186 $\pm$ 0.0015 & $-$0.0931 $\pm$ 0.0036 & $-$0.0158 $\pm$ 0.0013 \\
                $\alpha$ & 0.0008 $\pm$ 0.0004 & $-$0.0448 $\pm$ 0.0024 & $-$0.0008 $\pm$ 0.0004 \\
                $\alpha_\text{h}$ & 0.0022 $\pm$ 0.0006 & $-$0.0479 $\pm$ 0.0023 & 0.0011 $\pm$ 0.0002 \\
                $\alpha_\text{ex}$ & $-$0.0036 $\pm$ 0.0006 & $-$0.0816 $\pm$ 0.0029 & $-$0.0061 $\pm$ 0.0008 \\
                \hline
                {[$\alpha_\text{h}$/$\alpha_\text{ex}$]} & 0.0047 $\pm$ 0.0005 & ... & ... \\
    		\hline
	\end{tabular}
\end{table}

\subsubsection{Odd-Z Elements}
\label{ssec:lmc_oddz_over}

It has long been established that elements with odd atomic number are always more rare than elements with even atomic numbers except for hydrogen \citep{oddo1914molekularstruktur,harkins1917evolution}. 
The odd-Z elements that are explored in this work include Na, Al, and K. The odd-Z elements are mostly created by exploding massive stars. 



There is a large diversity in the gradients for the odd-Z elements (see Table \ref{tab:lmc_oddz_over_grad}). The [X/Mg] gradients are the flattest compared to the [X/Fe] and [X/H]. As expected for other groups of elements, the [X/Mg] gradients are more similar to the [X/Fe] gradients than the [X/H] gradients. Interestingly, the [Al/H] gradient is very steep versus any other odd-Z gradient, regardless of the fiducial element, with a value of $-$0.0682 $\pm$ 0.0036 dex/kpc. 

\begin{table}
	\centering
	\caption{A table of overall gradients for the odd-Z elements.}
	\label{tab:lmc_oddz_over_grad}
	\begin{tabular}{cccc}
		\hline
		Element & $\nabla_R$ [X/Fe] & $\nabla_R$ [X/H] & $\nabla_R$ [X/Mg]\\
                    & (dex/kpc) & (dex/kpc) & (dex/kpc)  \\
    	\hline
    	Na & 0.0017 $\pm$ 0.0019 & $-$0.0527 $\pm$ 0.0021 & $-$0.0002 $\pm$ 0.0019 \\
            Al & $-$0.0197 $\pm$ 0.0012 & $-$0.0682 $\pm$ 0.0036 & $-$0.0120 $\pm$ 0.0012 \\
            K & 0.0075 $\pm$ 0.0009 & $-$0.0412 $\pm$ 0.0026 & 0.0046 $\pm$ 0.0009 \\
    	\hline
	\end{tabular}
\end{table}

\subsubsection{Iron Peak Elements}
\label{ssec:lmc_ironpeak_over}

Cr, Mn, Fe, Co, Ni, and V are the iron peak elements investigated in this study. These elements typically enrich the ISM through Type Ia SNe, though some enrichment also does happen with Type II SNe \citep{kobayashi2006chemical}.  Fe tends to be more robustly measured than many other elements, with many good lines in optical stellar spectra. Therefore, it is often used as the main fiducial element used to track the overall ``metallicity'' of a star.



Comparing the three different fiducials, in general the gradients become large for increasing atomic number until Mn or Fe, where the gradients become shallower (see Table \ref{tab:lmc_iron_over_grad}). As with the other groups of elements, the [X/Mg] elements are in between the [X/Fe] and [X/H] gradients. The [Mn/H] is the steepest gradient out of all of the elements with a value of $-$0.1024 $\pm$ 0.0040 dex/kpc. 

\begin{table}
	\centering
	\caption{A table of overall gradients for the iron peak elements.}
	\label{tab:lmc_iron_over_grad}
	\begin{tabular}{cccc}
		\hline
		      Element & $\nabla_R$ [X/Fe] & $\nabla_R$ [X/H] & $\nabla_R$ [X/Mg]\\
                    & (dex/kpc) & (dex/kpc) & (dex/kpc)  \\
    		\hline
                V & $-$0.0079 $\pm$ 0.0012 & $-$0.0466 $\pm$ 0.0028 & $-$0.0105 $\pm$ 0.0013\\
                Cr & $-$0.0032 $\pm$ 0.0011 & $-$0.0666 $\pm$ 0.0024 & $-$0.0077 $\pm$ 0.0014\\
                Mn & $-$0.0247 $\pm$ 0.0009 & $-$0.1024 $\pm$ 0.0040 & $-$0.0228 $\pm$ 0.0013\\
                Fe & ... & $-$0.0380 $\pm$ 0.0022 & $-$0.0862 $\pm$ 0.0042\\
                Co & $-$0.0081 $\pm$ 0.0009 & $-$0.0528 $\pm$ 0.0023 & $-$0.0075 $\pm$ 0.0000\\
                Ni & 0.0010 $\pm$ 0.0004 & $-$0.0445 $\pm$ 0.0022 & $-$0.0013 $\pm$ 0.0000\\
    		\hline
	\end{tabular}
\end{table}

\subsubsection{Neutron Capture Elements}
\label{ssec:lmc_neutron_over}

Neutron capture elements are created through two processes: the rapid process (\textit{r}-process) and the slow process (\textit{s}-process). 
The $s$-process takes place inside massive stars in late stages of their evolution by the slow absorption of neutrons by heavy atoms, while the $r$-process happens in supernovae or binary neutron star mergers on the timescale of seconds.
In this work, we explore two neutron capture elements, namely Ce and Nd. These elements are created through both the \textit{r}- and \textit{s}-processes, though Ce prefers the \textit{s}-process much more than Nd does \citep{prantzos2020srprocess}. Note that there are reliable Ce and Nd abundance values only for a fraction of our LMC RGB sample, as previously mentioned (in section \ref{ssec:lmc_bacchus}). For Ce there are 486 stars with good values, and for Nd there are 386 stars with good values.

For the LMC, we find that the Nd gradient is always flatter than the Ce gradient for all three fiducials (see Table \ref{tab:lmc_cerium_over_grad}). 

\begin{table}
	\centering
	\caption{A table of overall gradients for the neutron capture elements.}
	\label{tab:lmc_cerium_over_grad}
	\begin{tabular}{cccc}
		\hline
		      Element & $\nabla_R$ [X/Fe] & $\nabla_R$ [X/H] & $\nabla_R$ [X/Mg]\\
                    & (dex/kpc) & (dex/kpc) & (dex/kpc)  \\
    		\hline
    		{[Ce/Fe]} & $-$0.0091 $\pm$ 0.0042 & $-$0.0642 $\pm$ 0.0078 & $-$0.006 $\pm$ 0.0039 \\
                {[Nd/Fe]} & 0.0054 $\pm$ 0.0038 & $-$0.0259 $\pm$ 0.0053 & 0.0012 $\pm$ 0.0046 \\
    		\hline
	\end{tabular}
\end{table}

\subsection{The Evolution of the LMC Abundance Ratio Gradients}\label{ssec:lmc_evolve}

Radial abundance gradients are sensitive to the particular formation history of a galaxy \citep[e.g.,][]{pagel1981abundances}.  If star formation proceeds at a higher rate at one radius compared to another, then this will be detectable in the radial abundances gradients.  With the addition of stellar age information (see Section \ref{ssec:lmc_age}), it is possible to track the evolution of the abundance gradients over time. The values for the [X/Fe], [X/H], and [X/Mg] for each of the age bins are tabulated in Tables \ref{tab:lmc_evolve_grad_fe}, \ref{tab:lmc_evolve_grad_h}, and \ref{tab:lmc_evolve_grad_mg} respectively. Here we measure evolution running from the oldest stars to the most recent. However, it is worth noting that stars move away from their birth location over time via radial migration \citep[e.g.,][]{sellwood2002radial}. This process flattens the abundance gradient over time \citep[e.g.,][]{minchev2013disk}. 
As a result, even though the gradients in mono-age populations contain memories of the birth gradients and thus reflect some information of the birth environment of the stars, gradients in mono-age populations have been modulated and generally smoothed out by radial migration.

\begin{figure*}
    \centering
    \includegraphics[width=\textwidth]{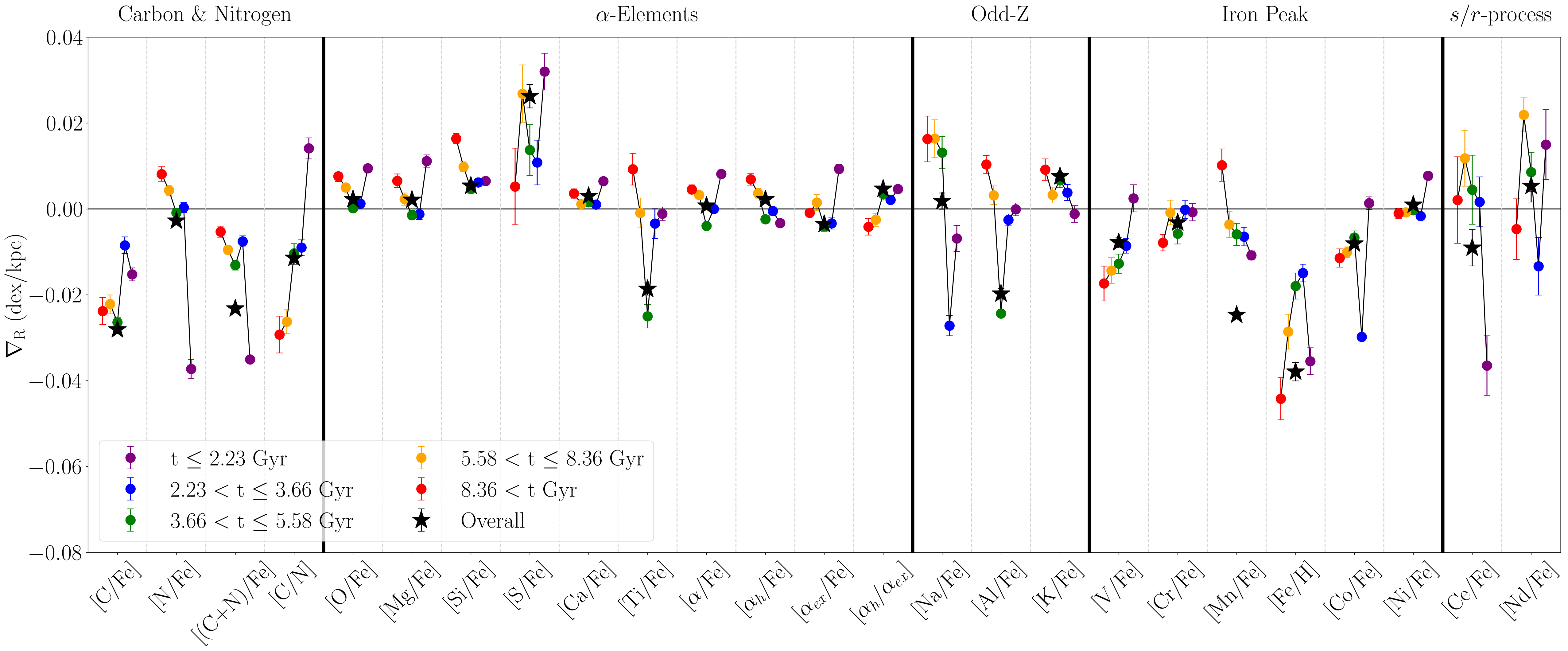}
    \caption{A figure showing each of the calculated [X/Fe] gradients. Also included here are [Fe/H], [C/N], [$\alpha_\text{h}$/Fe], [$\alpha_\text{ex}$/Fe] and [$\alpha_\text{h}$/$\alpha_\text{ex}$]. The abundances have been broken up into groups of nucleosynthetically similar elements with atomic number generally increasing to the right. The color of each dot corresponds to the different age bins for which a gradient was calculated. Black lines connecting each age-computed gradient have also been added to make their trends more clear. The error bars included are the MAD of the posterior distributions. The error bars tend to be small and this is due to using the MAD of the median abundance value of radial bins as a weighting factor instead of the dispersion of the whole bin.
    The value of each gradient here can be seen in Table \ref{tab:lmc_evolve_grad_fe}.}
    \label{fig:lmc_all_grads_fe}
\end{figure*}

\begin{figure*}
    \centering
    \includegraphics[width=\textwidth]{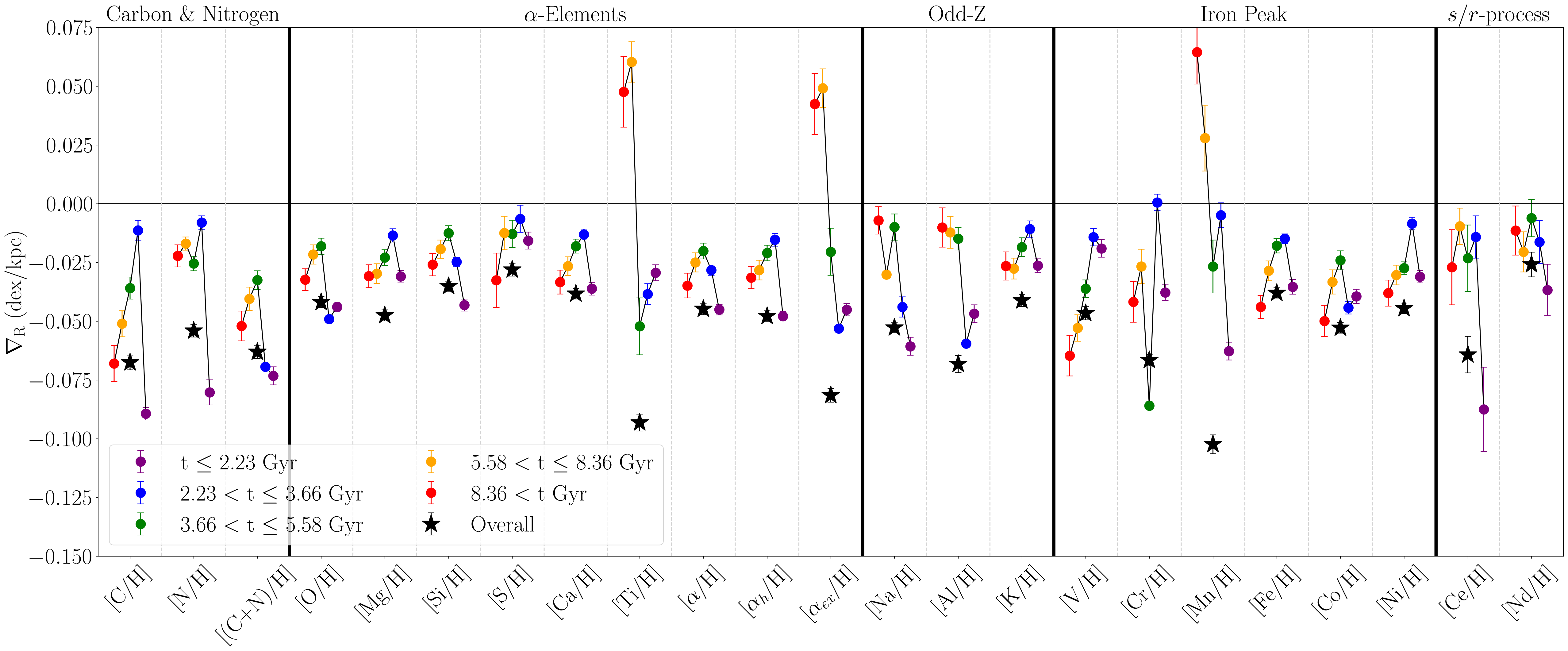}
    \caption{Same as Figure \ref{fig:lmc_all_grads_fe}, but for the [X/H] gradients. The [Fe/H] has also been included here because it has a hydrogen fiducial. Note that the [C/N] and [$\alpha_\text{h}$/$\alpha_\text{ex}$] are not included here because their fiducial abundances are not hydrogen and they have been included in Figure \ref{fig:lmc_all_grads_fe}. The value of each gradient is given in Table \ref{tab:lmc_evolve_grad_h}.}
    \label{fig:lmc_all_grads_h}
\end{figure*}

\begin{figure*}
    \centering
    \includegraphics[width=\textwidth]{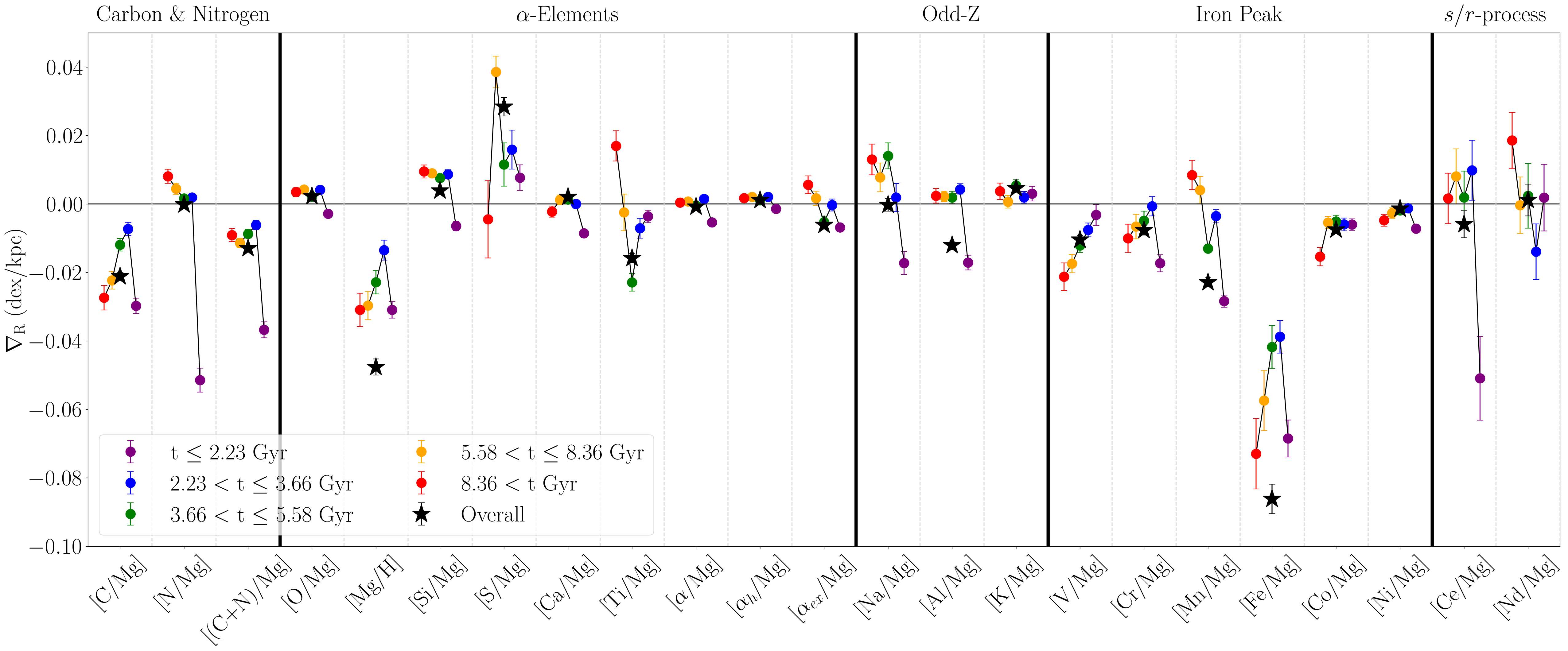}
    \caption{Same as Figure \ref{fig:lmc_all_grads_fe}, but for the [X/Mg] gradients. The [Mg/H] gradient has also been included here. The value of each gradient is given in Table \ref{tab:lmc_evolve_grad_mg}.}
    \label{fig:lmc_all_grads_mg}
\end{figure*}

For this section the most important points are summarized in Figures \ref{fig:lmc_all_grads_fe}, \ref{fig:lmc_all_grads_h}, \ref{fig:lmc_all_grads_mg}, which show the time dependence of all the measured gradients in this work. In general for each plot, atomic number increases from left to right and the elements have been grouped together with the previously mentioned groups. The abundance gradients for [C/N] and [$\alpha_\text{h}$/$\alpha_\text{ex}$] have been included in Figure \ref{fig:lmc_all_grads_fe}. A black line has been included to help guide the eye and make the shape of the trends more obvious with respect to time. The full radial trends can found in Figures \ref{fig:lmc_cn_radxfe}, \ref{fig:lmc_ind_alpha_radxfe}, \ref{fig:lmc_comb_alpha_radxfe}, \ref{fig:lmc_oddz_radxfe}, \ref{fig:lmc_ironpeak_radxfe}, and \ref{fig:lmc_oddz_radxfe}. The temporal behavior we measure for the gradients is described in more detail below.


\begin{table*}
	\centering
	\caption{A table of the gradients for each of the age bins. Horizontal lines have been added marking the division of the previously defined groups of elements. In general descending down the table corresponds to an increase in atomic number. Each age bin has its own separate column. This table also includes the evolution of the [Fe/H], [C/N] and [$\alpha_\text{h}$/$\alpha_\text{ex}$].}
	\label{tab:lmc_evolve_grad_fe}
	\begin{tabular}{cccccc}
        \hline
         & t $\leq$ 2.23 & 2.23 $<$ t $\leq$ 3.66 & 3.66 $<$ t $\leq$ 5.58 & 5.58 $<$ t $\leq$ 8.36 & 8.36 $<$ t \\
        Element & $\nabla_\text{R}$ & $\nabla_\text{R}$ & $\nabla_\text{R}$ & $\nabla_\text{R}$ & $\nabla_\text{R}$ \\
         & (dex/kpc) & (dex/kpc) & (dex/kpc) & (dex/kpc) & (dex/kpc) \\
        \hline 
        {[C/Fe]} & -0.0152 $\pm$ 0.0015 & $-$0.0085 $\pm$ 0.0020 & $-$0.0264 $\pm$ 0.0011 & $-$0.0222 $\pm$ 0.0022 & $-$0.0237 $\pm$ 0.0031 \\
        {[N/Fe]} & $-$0.0373 $\pm$ 0.0022 & 0.0003 $\pm$ 0.0011 & $-$0.0009 $\pm$ 0.0010 & 0.0044 $\pm$ 0.0012 & 0.0081 $\pm$ 0.0017 \\
        {[(C+N)/Fe]} & $-$0.0351 $\pm$ 0.0005 & $-$0.0076 $\pm$ 0.0013 & $-$0.0131 $\pm$ 0.0012 & $-$0.0095 $\pm$ 0.0010 & $-$0.0052 $\pm$ 0.0013 \\
        {[C/N]} & 0.0141 $\pm$ 0.0026 & $-$0.0090 $\pm$ 0.0018 & $-$0.0103 $\pm$ 0.0023 & $-$0.0262 $\pm$ 0.0028 & $-$0.0291 $\pm$ 0.0042 \\
        \hline
        {[O/Fe]} & 0.0095 $\pm$ 0.0010 & 0.0012 $\pm$ 0.0007 & 0.0001 $\pm$ 0.0008 & 0.0050 $\pm$ 0.0009 & 0.0076 $\pm$ 0.0012 \\
        {[Mg/Fe]} & 0.0111 $\pm$ 0.0014 & $-$0.0013 $\pm$ 0.0013 & $-$0.0015 $\pm$ 0.0010 & 0.0023 $\pm$ 0.0014 & 0.0065 $\pm$ 0.0016 \\
        {[Si/Fe]} & 0.0065 $\pm$ 0.0010 & 0.0062 $\pm$ 0.0010 & 0.0047 $\pm$ 0.0010 & 0.0099 $\pm$ 0.0011 & 0.0164 $\pm$ 0.0012 \\
        {[S/Fe]} & 0.0319 $\pm$ 0.0043 & 0.0108 $\pm$ 0.0051 & 0.0136 $\pm$ 0.0058 & 0.0270 $\pm$ 0.0066 & 0.0053 $\pm$ 0.0088 \\
        {[Ca/Fe]} & 0.0065 $\pm$ 0.0007 & 0.0010 $\pm$ 0.0006 & 0.0016 $\pm$ 0.0007 & 0.0012 $\pm$ 0.0009 & 0.0035 $\pm$ 0.0010 \\
        {[Ti/Fe]} & $-$0.0011 $\pm$ 0.0017 & $-$0.0035 $\pm$ 0.0036 & $-$0.0250 $\pm$ 0.0027 & $-$0.0009 $\pm$ 0.0034 & 0.0092 $\pm$ 0.0037 \\
        {[$\alpha$/Fe]} & 0.0081 $\pm$ 0.0009 & 0.000 $\pm$ 0.0006 & $-$0.0040 $\pm$ 0.0007 & 0.0032 $\pm$ 0.0008 & 0.0045 $\pm$ 0.0011 \\
        {[$\alpha_\text{h}$/Fe]} & $-$0.0033 $\pm$ 0.0005 & $-$0.0005 $\pm$ 0.0009 & $-$0.0025 $\pm$ 0.0008 & 0.0036 $\pm$ 0.0011 & 0.0069 $\pm$ 0.0013 \\
        {[$\alpha_\text{ex}$/Fe]} & 0.0093 $\pm$ 0.0010 & $-$0.0034 $\pm$ 0.0013 & $-$0.0041 $\pm$ 0.0012 & 0.0015 $\pm$ 0.0018 & $-$0.0009 $\pm$ 0.0006 \\
        {[$\alpha_\text{h}$/$\alpha_\text{ex}$]} & 0.0046 $\pm$ 0.0007 & 0.0021 $\pm$ 0.0009 & 0.0034 $\pm$ 0.0014 & $-$0.0026 $\pm$ 0.0015 & $-$0.0041 $\pm$ 0.0019 \\
        \hline
        {[Na/Fe]} & $-$0.0069 $\pm$ 0.0031 & $-$0.0271 $\pm$ 0.0024 & 0.0132 $\pm$ 0.0037 & 0.0166 $\pm$ 0.0045 & 0.0163 $\pm$ 0.0053 \\
        {[Al/Fe]} & $-$0.0001 $\pm$ 0.0015 & $-$0.0026 $\pm$ 0.0014 & $-$0.0244 $\pm$ 0.0009 & 0.0031 $\pm$ 0.0021 & 0.0103 $\pm$ 0.0021 \\
        {[K/Fe]} & $-$0.0012 $\pm$ 0.0019 & 0.0038 $\pm$ 0.0019 & 0.0068 $\pm$ 0.0017 & 0.0032 $\pm$ 0.0019 & 0.0092 $\pm$ 0.0024 \\
        \hline
        {[V/Fe]} & 0.0025 $\pm$ 0.0031 & $-$0.0086 $\pm$ 0.0017 & $-$0.0128 $\pm$ 0.0022 & $-$0.0143 $\pm$ 0.003 & $-$0.0174 $\pm$ 0.0041 \\
        {[Cr/Fe]} & $-$0.0008 $\pm$ 0.002 & $-$0.0002 $\pm$ 0.0021 & $-$0.0058 $\pm$ 0.0024 & $-$0.0008 $\pm$ 0.0029 & $-$0.0079 $\pm$ 0.0019 \\
        {[Mn/Fe]} & $-$0.0108 $\pm$ 0.0011 & $-$0.0064 $\pm$ 0.0021 & $-$0.0060 $\pm$ 0.0026 & $-$0.0037 $\pm$ 0.0030 & 0.0101 $\pm$ 0.0038 \\
        {[Fe/H]} & $-$0.0353 $\pm$ 0.0031 & $-$0.0150 $\pm$ 0.0020 & $-$0.0180 $\pm$ 0.0031 & $-$0.0286 $\pm$ 0.0042 & $-$0.0441 $\pm$ 0.0049 \\
        {[Co/Fe]} & 0.0013 $\pm$ 0.0015 & $-$0.0298 $\pm$ 0.0004 & $-$0.0067 $\pm$ 0.0015 & $-$0.0101 $\pm$ 0.0012 & $-$0.0114 $\pm$ 0.0022 \\
        {[Ni/Fe}] & 0.0077 $\pm$ 0.0009 & $-$0.0017 $\pm$ 0.0008 & $-$0.0003 $\pm$ 0.0007 & $-$0.0009 $\pm$ 0.0009 & $-$0.0011 $\pm$ 0.0011 \\
        \hline
        {[Ce/Fe]} & $-$0.0367 $\pm$ 0.0069 & 0.0017 $\pm$ 0.0058 & 0.0047 $\pm$ 0.0085 & 0.0121 $\pm$ 0.0065 & 0.0026 $\pm$ 0.0099 \\
        {[Nd/Fe]} & 0.0150 $\pm$ 0.0080 & $-$0.0133 $\pm$ 0.0068 & 0.0079 $\pm$ 0.0045 & 0.0220 $\pm$ 0.0039 & $-$0.0046 $\pm$ 0.0070 \\
    \end{tabular}
\end{table*}


\subsubsection{Carbon \& Nitrogen}
\label{ssec:lmc_cn_evolve}

\begin{figure*}
    \centering
    \includegraphics[width=0.95\textwidth]{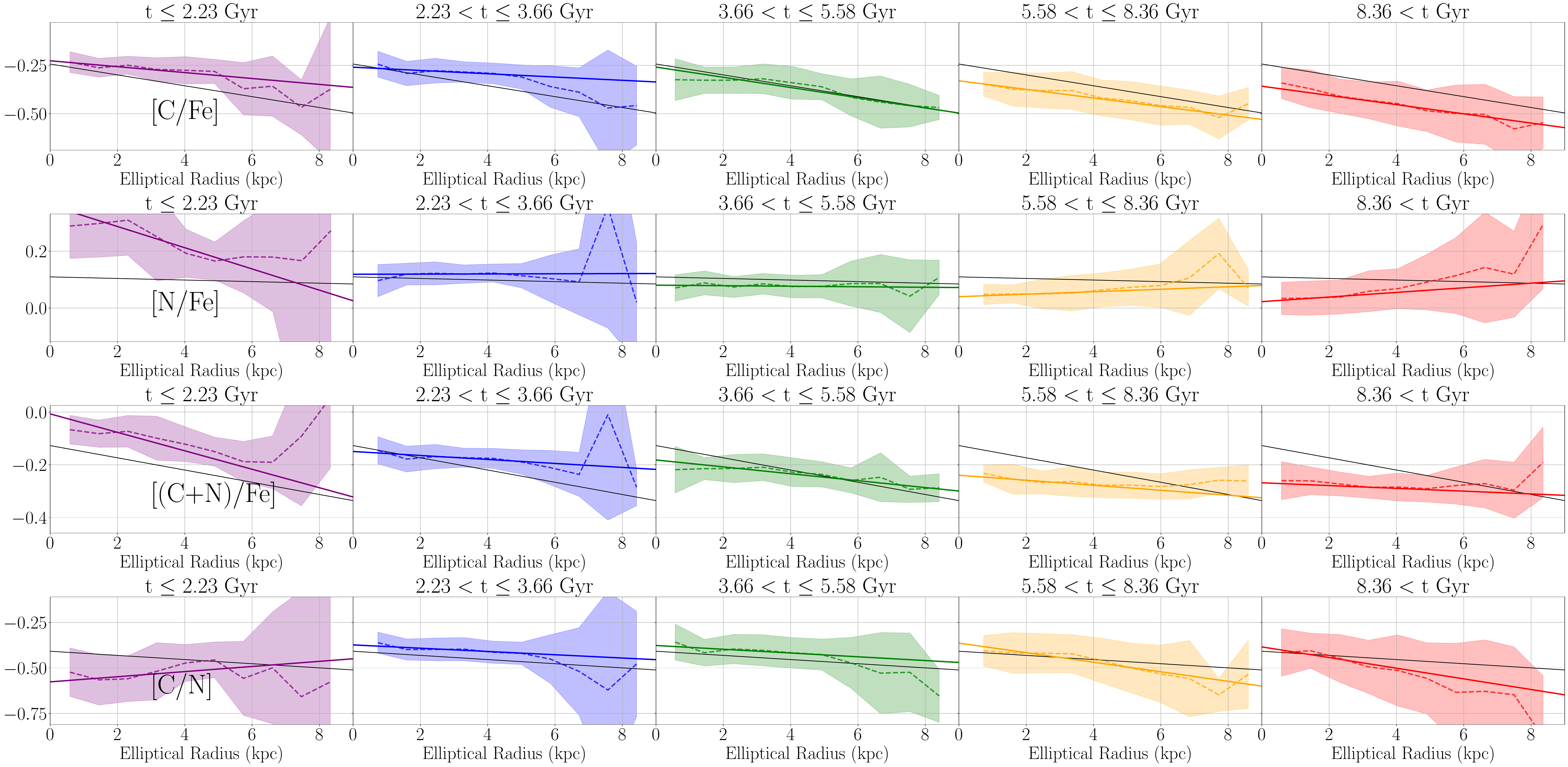}
    \caption{The radial abundance trends for the measured C and N abundances in the LMC. The particular abundance can be seen in the left-most panel of each row. For each row, age increases from left to right with colored lines representing the linear model fits to the binned values. A correspondingly colored dashed line represents the binned median abundance value with the band showing the dispersion. A black line in a panel shows the overall radial abundance trend without age binning and functions as a fiducial here. These plots show that the various C and N abundance ratio combinations increase over time in the Galaxy, with the trends being below the fiducial for older age bins and above for younger age bins. Also the [C/N] gradient clearly inverts for the youngest age bin. In an effort to capture the overall differences among panels, the range shown for each row is based on the fiducial line, even though this does result in some trends being cut off.}
    \label{fig:lmc_cn_radxfe}
\end{figure*}

The evolution in the C and N abundances and abundance ratios is varied, and sometimes contrasting. For the [X/Fe] gradients, [C/Fe] flattens out for younger stars compared to older ones showing an upward trend  (see Figure \ref{fig:lmc_all_grads_fe}). This contrasts with [N/Fe], which starts out positive, then flattens, and finally becomes negative over time. Figure \ref{fig:lmc_cn_radxfe} shows that the central [C/Fe] is always deficient compared to the Sun while steadily increasing from $\sim-$0.4 to $\lesssim-$0.2. Unlike what is seen in [C/Fe], the central [N/Fe] ratio starts close to solar, but increases around 5.6 Gyr ago to about $+$0.1 dex and then undergoes a major increase around 2.2 Gyr ago.



Combining the C and N abundances into ratios of [(C+N)/Fe] and [C/N] produces interesting results. The [(C+N)/Fe] gradient has a similar downward trend analogous to what is seen in [N/Fe], but there is a larger jaggedness (see Figure \ref{fig:lmc_all_grads_fe}). In terms of 
the actual abundance values, much like [N/Fe], there is a large increase in the central [(C+N)/Fe] value, but this is much less abrupt, starting between 3.7 and 5.6 Gyr ago and gradually increasing up to the present day (see Figure \ref{fig:lmc_cn_radxfe}). 
In contrast, [C/N] shows a different evolutionary behavior.
The [C/N] gradient changes almost linearly from $-$0.03 at old ages to $+$0.015 at the youngest ages.  We can understand this change by remembering that [C/N] is sensitive to age for RGB stars, because the dredge-up of material is sensitive to mass as mentioned in \cite{Hasselquist2020} and references therein. 
The age-behavior for metal-poor RGB stars has not been characterized yet.  One possible interpretation of the age dependent behavior of the [C/N] gradient, is that it is merely a result of the metallicity evolution at different radii.   However, the [Fe/H] gradient does not follow this pattern, as can be seen in Figure \ref{fig:lmc_all_grads_h}.  Instead, the [C/N] is likely the result of the changing age distribution at different radii.  Old stars exist at essentially all radii in the LMC, while the youngest stars are more centrally concentrated because that is where present day star formation is occurring. 


The [C/H], [N/H], and [(C+N)/H] abundance gradients all show a peculiar trend in their evolution. Each starts steep, but flattens out over time until just after 2.2 Gyr ago when the gradients then become steeper again (Fig.~\ref{fig:lmc_all_grads_h}). This U-shaped time trend is interesting as the extreme point matches in each trend (see Figure \ref{fig:lmc_all_grads_h}). The [C/Mg] gradient also shares this behaviour, but the [N/Mg] and [(C+N)/Mg] gradients do not (Fig.~\ref{fig:lmc_all_grads_mg}). 


\subsubsection{\texorpdfstring{$\alpha$}\,-Elements}
\label{ssec:lmc_alpha_evolve}


Many of the [X/Fe] gradients for the individual $\alpha$-elements as well as two of the composite $\alpha$ abundance ratios ([$\alpha$/Fe] and [$\alpha_\text{ex}$/Fe]) show a curious U-shaped trend (see Figure \ref{fig:lmc_all_grads_fe}). Abundance ratios that show a flattening positive gradient and then more recent steepening include [O/Fe], [Mg/Fe], [Ca/Fe], [Ti/Fe], [$\alpha$/Fe], and [$\alpha_\text{ex}$/Fe]. [S/Fe] may appear to potentially show this trend also, but it is noisy. To a lesser extent, signs of this trend are seen in [Si/Fe], where its gradient flattens but then stays around $+$0.0060 dex/kpc. The [$\alpha_\text{h}$/Fe] gradient shows a downward trend and inverts between 5.6 and 3.7 Gyr ago. For the most part, Figures \ref{fig:lmc_ind_alpha_radxfe} and \ref{fig:lmc_comb_alpha_radxfe} show that the radial trends evolve much in the same way regardless of the abundance.

By and large, the [X/H] gradients for the $\alpha$-element abundances  show similar evolutionary trends: All but [Ti/H] and [$\alpha_\text{ex}$/H] show an inverted U-shaped trend with an extreme point around $\lesssim$2.2 Gyr (see Figure \ref{fig:lmc_all_grads_fe}). Unlike the [X/Fe] gradients, there is much more consistency across different elements. 

The [O/Mg], [Si/Mg], [$\alpha$/Mg], and [$\alpha_\text{h}$/Mg], as well as potentially [Ca/Mg], gradients all evolve in a similar manner. For most of the history of the LMC, these gradients did not change until 2.2 Gyr ago. All of these elements, except [Ti/Mg], have very slightly positive gradients, but the youngest stars have slightly negative ones.

\subsubsection{Odd-Z Elements}
\label{ssec:lmc_oddz_evolve}



The evolution of the [Na/Fe] and [Al/Fe] gradients are quite interesting because they both start out positive but become negative and finally tend to 0.0 dex/kpc (see Figure \ref{fig:lmc_all_grads_fe}). Initially, the [Na/Fe] gradients do not evolve until more recently than 3.36 Gyr ago. Unlike [Na/Fe], the [Al/Fe] gradient immediately starts to flatten, going from $+$0.0103 $\pm$ 0.0021 dex/kpc to $+$0.0031 $\pm$ 0.0021 dex/kpc. In fact, it appears that the evolution in the [Na/Fe] gradient lags the analogous evolution in the [Al/Fe] gradient. This is somewhat reflected in the radial trends in Figure \ref{fig:lmc_oddz_radxfe}. The shows more chaotic behavior compared to [Al/Fe], but this could be because [K/Fe] is less reliably measured by APOGEE. 
The radial trend for [K/Fe] shows also that the central abundance does not change much over time.

As for the [X/H] gradients of the odd-Z elements, [K/H] evolves much like what is seen for the C and N gradients as well as the $\alpha$-element gradients (see Figure \ref{fig:lmc_all_grads_h}). The evolutionary trend for [Na/H] has some jaggedness where the gradient for ages between 5.58 and 8.36 Gyr is steep with a value of $-$0.0301 $\pm$ 0.0015 dex/kpc, while the oldest stars and stars ages 3.66--5.58 Gyr have gradients of $-$0.0071 $\pm$ 0.0058 dex/kpc and $-$0.0100 $\pm$ 0.0056 dex/kpc, respectively. The youngest gradients for [Na/H] are then steeper and more negative than any of these values. The [Al/H] gradient does not experience much evolution until about 5.6 to 3.7 Gyr ago when there is an abrupt steepening in the gradient with a slight upturn for stars younger than 2.2 Gyr.

The [Na/Mg] gradient evolution  (Figure \ref{fig:lmc_all_grads_mg}) resembles that of [Na/H] (Figure \ref{fig:lmc_all_grads_h}) in its shape, but the latter is offset from the former by being $\sim$0.2 dex/kpc lower. On the other hand, the [Al/Mg] gradient stays almost flat for the entire history of the LMC until the present age bin where the gradient drops precipitously to $-$0.0172 $\pm$ 0.0021 dex/kpc. The [K/Mg] gradient shows even less evolution than [Al/Mg], by remaining flat even for the youngest ages.
  

\subsubsection{Iron Peak Elements}
\label{ssec:lmc_ironpeak_evolve}

The iron peak [X/Fe] abundances show varied evolution in their gradients (see Figure \ref{fig:lmc_all_grads_fe}). The [V/Fe] and [Cr/Fe] both show an upward trend, though this is more prominent for [V/Fe]. Interestingly, the radial trends of these elements reveal that there is not much change in the central abundance of the LMC (Figure \ref{fig:lmc_ironpeak_radxfe}). The [Mn/Fe] gradient stands out among the iron peak [X/Fe] gradients as it is the only one that has a continuous downward trend over all age bins. The [Mn/Fe] gradient starts at $+$0.0101 $\pm$ 0.0038 dex/kpc for the oldest age, but then in the next age bin the gradient has immediately dropped changing sign, becoming $-$0.0037 $\pm$ 0.0030 dex/kpc. From there [Mn/Fe] continues to become even more negative over time. The radial trend for [Mn/Fe] displayed in Figure \ref{fig:lmc_ironpeak_radxfe} shows a large increase in the central LMC abundance between the oldest and second oldest age bins, but then the ratio increases less drastically afterwards, corresponding to what is seen in the gradient. Unlike any other [X/Fe] gradient, the overall gradient without age binning for [Mn/Fe] falls outside the range suggested by the evolutionary trend of the gradient. The cause of this behavior in the global gradient is not clear (see Figure \ref{fig:lmc_all_grads_fe}), but is most likely due to some effects caused by the binning. The evolution of the [Co/Fe] gradient is also quite unique with a clear increasing trend, but the gradient for the ages between 2.2--3.7 Gyr has an anomalous value of $-$0.0298 $\pm$ 0.0004 dex/kpc. Another possibility is the [Co/Fe] also shows the U-shaped trend share among other elements.

Similarly to many other groups of elements, the majority of the [X/H] gradients for the iron peak show an inverted U-shaped with an extreme point at the same time $\sim$2.23$-$3.66 Gyr ago (Figure \ref{fig:lmc_all_grads_h}). [Cr/H] and [Mn/H] are the only abundances that do not show this trend, though [Mn/H] deviates more than [Cr/H] from the ``normal'' pattern. As with the overall [Mn/Fe] gradient, the overall [Mn/H] gradient is unusually negative with a value of $-$0.1024 $\pm$ 0.0040 dex/kpc, which is $\sim$0.0400 dex/kpc from the closet gradient when binning by age.


The [X/Mg] ratios for both Cr and Fe 
show the previously mentioned inverted U-shaped trend while the other iron peak elements do not (see Figure \ref{fig:lmc_all_grads_mg}). The [V/Mg] gradient shows a flattening over time, becoming shallower. The [Mn/Mg] gradient shows a similar pattern, albeit scaled version, to the trend in [Mn/H]. The [Mn/Mg] gradient initially shows a downward trend until 2.23--3.66 Gyr ago, when it suddenly flattens before becoming relatively steep for the youngest stars, with a value of $-$0.0284 $\pm$ 0.0018 dex/kpc. The [Co/Mg] gradient starts at $-$0.0154 $\pm$ 0.0027 dex/kpc and then increases to $\sim$ $-$0.0060 dex/kpc up to the present. Conversely, the [Ni/Mg] gradient shows little evolution with no significant change, remaining at $\sim$ $-$0.0020 dex/kpc and then dropping to $-$0.0072 $\pm$ 0.0012 dex/kpc for young ages.

\subsubsection{Neutron Capture Elements}
\label{ssec:lmc_neutron_evolve}

For the neutron capture elements, there are not as many stars with well-measured abundances. The [Ce/Fe] gradient trend shows an inverted U-shape while this is not the case for the [Nd/Fe] gradient  (Figure \ref{fig:lmc_all_grads_fe}). For [Ce/Fe], the only gradient that decisively deviates from the general trend relative to the rest of the [Ce/Fe] gradients is the very steep young gradient with a value of $-$0.0367 $\pm$ 0.0069 dex/kpc. There is a delay in the increase of the central [Ce/Fe] value, which only slightly changes until the most recent ages where there is a sharp increase (Figure \ref{fig:lmc_sr_radxfe}). The [Nd/Fe] gradient evolution is similar to that for [S/Fe], but shifted to lower values. 

Both the [Ce/H] and [Nd/H] abundance gradients show signs of the inverted U-shaped evolution present in the [X/H] gradients of almost all the elements (Figure \ref{fig:lmc_all_grads_h}). However, this is not the case for the [X/Mg] gradients (Figure \ref{fig:lmc_all_grads_mg}, where the [Ce/Mg] gradient stays around $+$0.0015--0.0020 dex/kpc and then becomes quite steep at $-$0.0507 $\pm$ 0.0123 dex/kpc. Meanwhile, the [Nd/Mg] gradient does show what appears to be a U-shaped trend with an extreme point  between 2.23 and 3.66 Gyr ago as seen in other abundance gradients. 





\subsection{Age-[X/Fe] Trends Within Radial Bins}
\label{ssec:lmc_age_xfe_trends}

So far we have discussed the behavior of radial abundance gradients and their temporal evolution.  However, with our temporal information we can directly investigate how abundances change with as a function of age across the entire galaxy and also how they they compare when they are separated into different spatial regions.  As shown in Figure \ref{fig:lmc_annuli_axfe_relations}, we break the spatial coverage into three radial zones and then into a North and South region for a total of six spatial zones (see Section \ref{ssec:lmc_age_xfe_trend}). Also when discussing over or under abundance in the section, this will be relative to a fiducial trend that was found without any spatial binning.





\subsubsection{Carbon \& Nitrogen}
\label{sssec:lmc_cn_age_xfe_trends}

\begin{figure*}
    \centering
    \includegraphics[width=0.85\textwidth]{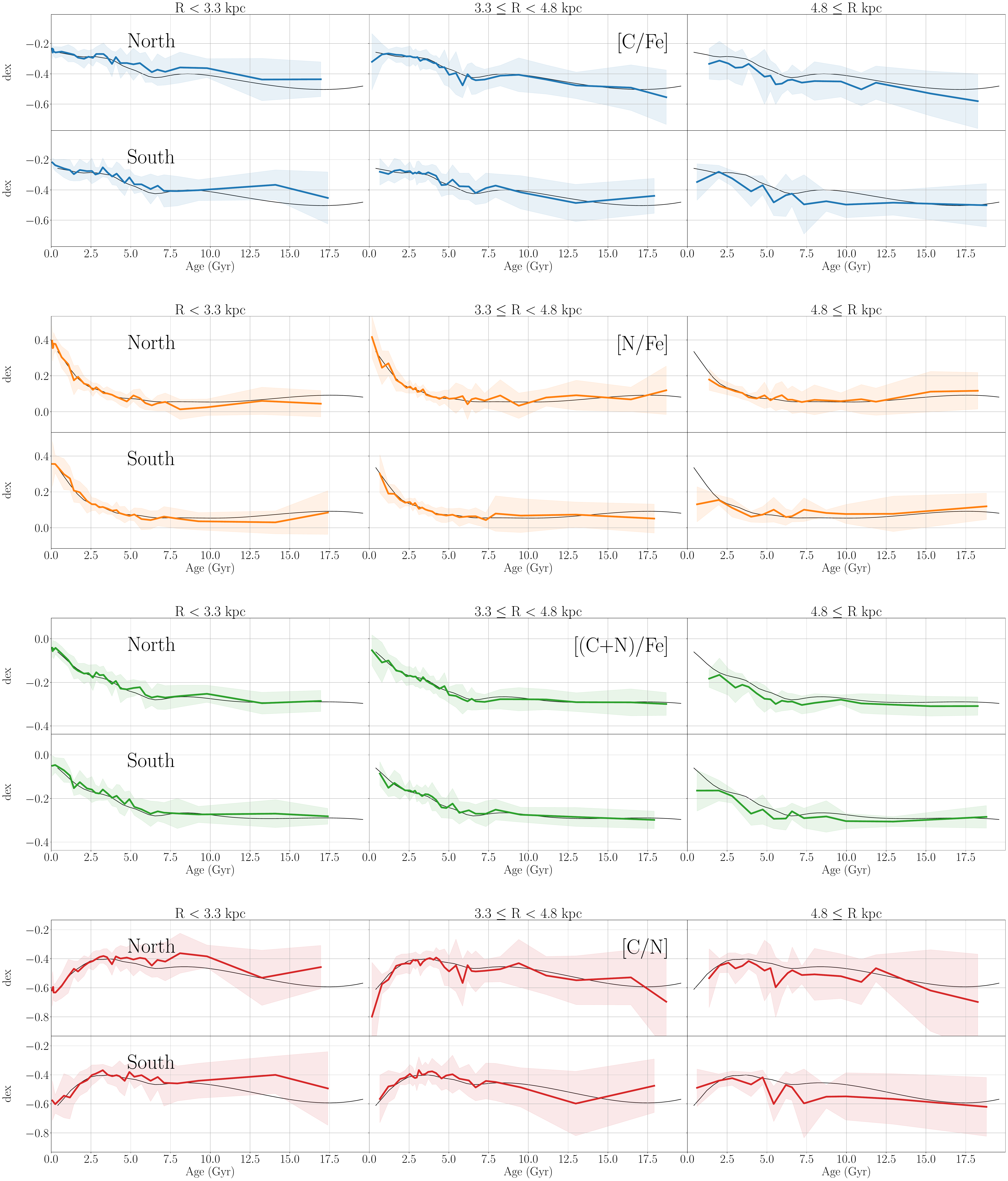}
    \caption{The LMC age-[X/Fe] trends for C and N in three radial bins for each of the North and South halves. The continuous spectrum of ages is based on the age binning outlined in Section \ref{ssec:lmc_age_xfe_trend}. The top row in each panel shows the age-[X/Fe] trends for the northern part of the LMC and the bottom row shows the trends in the south. The LMC is split into three radial bins that contain an equal number of stars. For each sub-panel the age-[X/Fe] trend (red line) is included with a (pink) band of width 1$\sigma$.   The black fiducial line shows the age-[X/Fe] trend without any binning. Clearly the [N/Fe] and [(C+N)/Fe] show a sharp increase in recent times.}
    \label{fig:lmc_cn_axfe}
\end{figure*}

Figure \ref{fig:lmc_cn_axfe} shows the age-[X/Fe] trends for [C/Fe], [N/Fe], and [(C+N)/Fe] separated into the six spatial zones. 
The age-[X/Fe] trends for [C/Fe], [N/Fe], and [(C+N)/Fe] are relatively flat, especially [N/Fe] and [(C+N)/Fe], up to 5 Gyr ago. For radii beyond 4.8 kpc, the LMC appears to be deficient in these abundances for both the north and south, although in the south, stars older than 12 Gyr follow the fiducial age-trend. The inner galaxy, on the other hand, is overabundant in these elements, especially for ages greater than 5 Gyr, while intermediate radii follow the fiducial age-trend quite well. The age-[N/Fe] trend follows the fiducial very well throughout the galaxy and appears to exponentially increase starting between 2 and 5 Gyr ago. Much like [N/Fe], [(C+N)/Fe] is nearly constant for the oldest ages and is deficient in the outer galaxy, much like [C/Fe]. The increase seen within the last 5 Gyr is quite linear similar to [C/Fe] but unlike [N/Fe].

The age-[C/N] trend is mostly flat throughout the LMC. The outer radial region shows that the periphery is low in [C/N] while the inner galaxy is rich in [C/N] except for the youngest stars with ages less than 2.5 Gyr. There is also a prominent downturn for young ages for the inner galaxy and especially for $3.3 < R \leq 4.8$ kpc.





\subsubsection{\texorpdfstring{$\alpha$}\,-Elements}
\label{sssec:lmc_alpha_age_xfe_trends}



Most of the age-[X/Fe] trends for the $\alpha$-elements do not show much evolution (see Figure \ref{fig:lmc_ind_alpha_axfe}). Even with how little some of the age-[X/Fe] trends evolve, there is an interesting feature shared by many of these elements ([O/Fe], [Mg/Fe], [Si/Fe], [Ca/Fe], and [Ti/Fe]) is a hump-like feature for younger stars. Each of the elements show an increase in the [X/Fe] abundance ratio starting around 5.0$-$7.5 Gyr ago with a peak $\lesssim$ 2.5 Gyr ago followed by a drop off for the youngest stars. The only element for which this is not seen in the individual $\alpha$-elements is [S/Fe], which remains mostly constant with time. For the inner radii, stars older than about 7.5 Gyr are deficient in [O/Fe], [Si/Fe], and slightly in [Ti/Fe] as these trends fall below the fiducial trend. The outer LMC in the south is especially overabundant in [O/Fe] and the outer LMC is [Si/Fe]-rich.


The composite $\alpha$-element abundance ratios as a function of age look very similar to those made from individual elements (Figure \ref{fig:lmc_comb_alpha_axfe}), which is not unexpected. The ratios [$\alpha$/Fe], [$\alpha_\text{h}$/Fe], and [$\alpha_\text{ex}$/Fe] all show the hump seen in the individual $\alpha$-elements. The [$\alpha_\text{h}$/$\alpha_\text{ex}$] age-[X/Fe] trend evolves much like the [S/Fe] as it is flat over the lifetime of the LMC and there is potentially a slight downturn close to the present day.

\subsubsection{Odd-Z Elements}
\label{sssec:lmc_oddz_age_xfe_trends}

The age-[Na/Fe] trend is mostly flat suggesting [Na/Fe] production has been constant (Figure \ref{fig:lmc_oddz_axfe}). Curiously, the outer LMC is [Na/Fe]-rich as the trends in both the north and south show, though this is less true for more recent times. For $R < 4.8$ kpc, the age-[Na/Fe] trend follows the fiducial in the south. Also, on average, the inner northern region seems to follow the fiducial.

Of the odd-Z elements, [Al/Fe] has the most dramatic change at younger ages, as seen in the middle panel of Figure \ref{fig:lmc_oddz_axfe}. Starting between 5.0 and 7.5 Gyr ago, the production of [Al/Fe] ramped up, with a change of almost $+$0.2 dex. Before this, the age-[Al/Fe] trend is relatively flat. Overall the age-[Al/Fe] trends throughout the LMC track the fiducial very well regardless of the bin. The individual trends have only minor deviations from the fiducial trend.


While [Na/Fe] has a flat age trend and [Al/Fe] is mostly constant with an increasing trend for more recent times, [K/Fe] is fairly constant for older ages, although it shows a hump within the last few Gyr (i.e., a slow increase with sudden drop at the youngest ages; see bottom panel of Figure \ref{fig:lmc_oddz_axfe}). For ages older than 5.0$-$7.5 Gyr, the age-abundance trend is flat and slightly overabundant, though younger stars appear to be somewhat deficient. The intermediate radii ($3.3 \leq R < 4.8$ kpc) for the north and south follow the fiducial trendline. The inner radii ($R < 3.3$ kpc) show a distinctive hump at young ages. 

\subsubsection{Iron Peak Elements}
\label{sssec:lmc_iron_age_xfe_trends}

The age-[X/Fe] trends for the iron-peak elements are very flat except for [Mn/Fe] and [Fe/H]. The subpopulations (different spatial bins) for  [Mn/Fe] tend to follow its fiducial trendline much more closely than [Fe/H], setting these two elements apart from each other. The outer galaxy is lacking in [Fe/H] for the north and the south as the trend lines clearly show the age-[Fe/H] trends fall below the average represented by the fiducial. More often than not, the other LMC positional regions are [Fe/H]-rich for older ages.  This makes sense as a general enrichment of Fe over time.

[Co/Fe] shows a hump like many of the other non-iron peak elements. There are some faint hints of a hump in other iron peak elements, but nowhere as strong as for Co. Like the other elements showing similar humps, it is the most prominent in the inner galaxy.

[V/Fe], [Cr/Fe] and [Ni/Fe] have flat trends, but a downturn at recent times that is strongest in the inner galaxy for [V/Fe] and [Ni/Fe], but not for [Cr/Fe]. [Cr/Fe] is flat and remains flat while also matching the fiducial trendline. The outer galaxy appears to be on average more poor in [V/Fe] compared to the LMC out to 4.8 kpc.

\subsubsection{Neutron Capture Elements}
\label{sssec:lmc_neutron_age_xfe_trends}

For the neutron capture elements, we only use two radial bins due to the low number of stars with good neutron capture abundance measurements. The first radial bin corresponds to the inner disk region with $R < 3.5$ kpc and the second radial bin includes everything outside of 3.5 kpc. The 3.5 kpc boundary was chosen based on \cite{choi2018reddening}, specifically, the fact that the inner disk corresponds to radii that are $\lesssim$ 3.5 kpc. 

The [Ce/Fe] ratio increases with time, though a dramatic change happens for stars born around 2.5 Gyr ago (see Figure \ref{fig:lmc_sr_axfe}). Generally, the age-[Ce/Fe] trend follows the fiducial line much most of the other elements, but there is a slightly higher deviation from the average trend within the last 5.0 Gyr or so. Since there is a low number of stars for [Ce/Fe] it is not clear if this is due to low number statistics or not.


Unlike [Ce/Fe], the age-[Nd/Fe] trend is mostly flat for all ages with no noticeable increase (see Figure \ref{fig:lmc_sr_axfe}). Compared to [Ce/Fe], there are even less stars with reliable values of [Nd/Fe], which is obvious when looking at the trends. Each of the northern bins has $\sim$130 stars, while the southern bins only have  $\sim$60 each. The age-[Nd/Fe] trends seem to follow the fiducial, but the inner southern field shows a [Nd/Fe] deficiency with respect to the fiducial age-[Nd/Fe] trendline.

\section{Discussion}
\label{sec:lmc_discussion}





Here we discuss the relationship between the abundance gradients, the age-[X/Fe] trends and the chemical evolution for the LMC. 


\subsection{LMC Metallicity Gradient}
\label{ssec:lmc_feh_grad_discuss}

\begin{figure}
    \centering
    \includegraphics[scale=0.3]{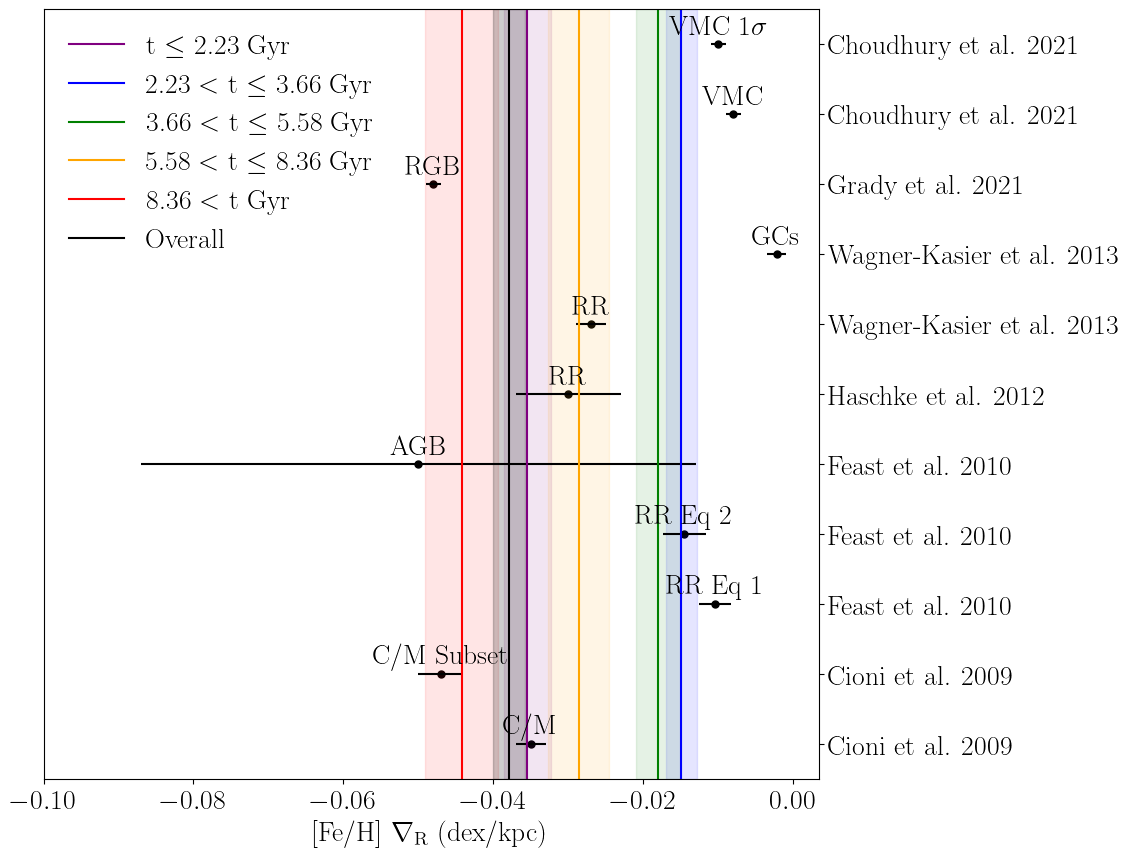}
    \caption{A comparison of the overall metallicity gradient found in this work to those from various previous studies. The vertical black line represents the overall metallicity gradient and the grey-black band represents the calculated error in the gradient. The gradient for each of the age bins is similarly shown. \protect\cite{cioni2009grad,feast2010lmcgrad,choudhury2021vmc} show agreement with the overall gradient, while other sources agree with the gradients found with different age bins in this work. In many cases, many of these sources calculate the gradient multiple ways. For more information see Section \ref{ssec:lmc_feh_grad_discuss} and the references therein.}
    \label{fig:lmc_lit_compare_feh}
\end{figure}

The LMC is found to have an overall radial metallicity (meaning [Fe/H]) gradient of $-$0.0380 $\pm$ 0.0022 dex/kpc. This value agrees with the value of $-$0.035 $\pm$ 0.0020 dex/kpc found using ratios of C- and M-type AGB stars \citep{cioni2009grad}, the value of $-$0.0500 $\pm$ 0.0370 dex/kpc found by \cite{feast2010lmcgrad} using photometric metallicities of a sample of AGB stars, and the value of $-$0.0300 $\pm$ 0.0070 dex/kpc found by \cite{haschke2012metallicity} also using photometric metallicities (see Table~\ref{tab:lmc_lit_grads} and Figure~\ref{fig:lmc_lit_compare_feh}). 


\cite{cioni2009grad} explores the metallicity gradients of the MCs and M33 using the C/M ratio. It was shown that the ratio of C- to M-type AGB stars in a region corresponds to the average [Fe/H] by \cite{battinelli2005cm_feh}. Calculating the [Fe/H] gradient in the LMC with the C/M ratio gives a value of $-$0.035 $\pm$ 0.0020 dex/kpc out to $\sim$10 kpc. This agrees quite well with the derived gradient in this work. By removing all [Fe/H] measurements for which $\sigma_\text{[Fe/H]} > 0.2$ gives a value of $-$0.0470 $\pm$ 0.0030 dex/kpc for the subset. This gradient does not overlap with the calculated overall gradient here, but it does agree with the gradient found for the oldest stars.

It has been shown that there is a relation between the period and [Fe/H] of RR Lyrae stars \citep[e.g.,][]{sandage1993rrlyrae,layden1995rrlyrae,sarajedini2006rrlyrae}.  \cite{feast2010lmcgrad} make use of two different [Fe/H]-Period relations to calculate the metallicity gradient in the LMC. Their dataset comes from OGLE \citep{soszynski2009rrlyrae} and extends out to 5 kpc in the LMC with stars brighter than $I = 
13.8$. Feast et al. also calculate the metallicity gradient for a sample of AGB stars. Only one of the gradients calculated for the [Fe/H]-Period relation overlaps with the calculated gradients here.
The gradient derived using Equation 2 in Feast et al. 
agrees with the youngest two [Fe/H] gradients here, but not the overall calculated one. The AGB gradient actually overlaps with the overall gradient as well as all of the gradients for each age bin, but the AGB gradient does have a large uncertainty. 

\cite{haschke2012metallicity} and \cite{wagner2013grad} also derive metallicity gradients using RR Lyrae and obtain similar values to each other. The \cite{haschke2012metallicity} gradient has a value of $-$0.0300 $\pm$ 0.0070 dex/kpc and the \cite{wagner2013grad} gradient of $-$0.0270 $\pm$ 0.02 dex/kpc, meaning these both match very well to the 5.58 $<$ age $\leq$ 8.36 Gyr [Fe/H] gradient of $-$0.0286 $\pm$ 0.0042 dex/kpc. The \cite{haschke2012metallicity} gradient also overlaps with the youngest [Fe/H] in addition to the overall gradient as previously mentioned. The globular cluster gradient in \cite{wagner2013grad} is much flatter than the one found with RR Lyrae stars (see Table~\ref{tab:lmc_lit_grads} and Figure~\ref{fig:lmc_lit_compare_feh}).

Another paper that calculates a metallicity gradient out to $\sim$12$^\circ$ for the LMC is \cite{grady2021magellanic}, who use photometric [Fe/H] values derived from machine learning with Gaia red giants. Their gradient is then validated using measured APOGEE [Fe/H]. It was found that their calculated [Fe/H] values fell within an RMSE of $\sim$0.15 dex of the APOGEE values. In the end they find that the LMC has a [Fe/H] gradient of $-$0.0480 $\pm$ 0.0010 dex/kpc, which definitely differs from our value of $-$0.0380 $\pm$ 0.0022 dex/kpc. While their overall method does differ from ours, it is possible that some of the difference arises from their APOGEE sample compared to ours. Their APOGEE sample is derived from APOGEE DR16 \citep{ahumada2020dr16} whereas the work herein used DR17. The APOGEE DR17 catalogue has effectively doubled the LMC coverage compared to DR16 and additionally some stars have improved [Fe/H] measurements in DR17. For more on the data and method used in the \cite{grady2021magellanic} study see Section 2 and 3 respectively of that paper. 

Another photometric study using the VMC data out to a radius of $\sim$6 kpc in the LMC by \citet{choudhury2021vmc} does not find radial gradients consistent with any calculated in our work. Using the total VMC LMC data and the slope of the RGB branch, these authors found a gradient of $-$0.0080 $\pm$ 0.0010 dex/kpc, and repeating the analysis for the 1$\sigma$ clipped [Fe/H]-RGB slope relation set they found a slightly different gradient of $-$0.0010 $\pm$ 0.0010 dex/kpc. 



It is clear that there is considerable variation amongst different calculated metallicity gradients for the LMC in the literature, even when different studies employ 
the same source tracers. This is also similar to the tension in derived MW abundance gradients \citep[e.g.,][]{braganca2019gradients}. With the high spectral resolution of the APOGEE survey as well as the number of observed stars and its coverage, it is certain that the metallicity gradients in this work has been calculated on a sample of data with some of the lowest uncertainties for individual stars suggesting high reliability.





\begin{table}
	\centering
	\caption{Table of literature values for the LMC [Fe/H] gradient. In the last column, RR Eq 1 and RR Eq 2 refer to two different period-[Fe/H] for RR Lyrae stars, which are Equations 1 and 2 in \protect\cite{feast2010lmcgrad} respectively. Also for \protect\cite{choudhury2021vmc} VMC corresponds to the gradient using the VMC survey data and VMC 1$\sigma$ is the gradient derived for the dataset using the 1$\sigma$-clipped VMC [Fe/H]-RGB slope relation. \protect\cite{grady2021magellanic} derives the gradient from [Fe/H] isochrones and calibrates them with APOGEE.}
	\label{tab:lmc_lit_grads}
	\begin{tabular}{cccc}
		\hline
		  Source & $\nabla_{\rm R}$ & $\nabla_{\rm R}$ Error & Method \\
                   & (dex/kpc) & (dex/kpc) & \\
		\hline
                Cioni et al. 2009 & $-$0.0350 & 0.0020 & C/M \\
                ---''--- & $-$0.0470 & 0.0030 & C/M Subset \\
                Feast et al. 2010 & $-$0.0104 & 0.0021 & RR Equation 1 \\
                ---''--- & $-$0.0145 & 0.0029 & RR Equation 2 \\
                ---''--- & $-$0.0500 & 0.0370 & AGB \\
                Haschke et al. 2012 & $-$0.0300 & 0.0070 & RR \\
                Wagner-Kasier et al. 2013 & $-$0.0270 & 0.0020 & RR \\
                ---''--- & $-$0.0022 & 0.0013 & GCs \\
                Grady et al. 2021 & $-$0.0480 & 0.0010 & RGB \\
                Choudhury et al. 2021 & $-$0.0080 & 0.0010 & VMC \\
                ---''--- & $-$0.0100 & 0.0010 & VMC 1$\sigma$  \\
		\hline
	\end{tabular}
\end{table}

\subsection{The Evolution of Abundance Gradients of the LMC}

We present here one of the first studies to explore the evolution of abundance gradients in the LMC. To do this the ages of 6130 field RGB stars were derived in Paper I. The stars were then placed into five different age bins for each of which an abundance gradient was found. While not included here, during the project analysis we investigated and determined that changing the binning scheme did not greatly affect the results other than minor differences. 








Many of the gradients show a U-shaped trends in their evolution with an extreme point between 2.33 and 3.66 Gyr ago. The gradients with this feature include [O/Fe], [Mg/Fe], [Ca/Fe], [Ti/Fe], [$\alpha$/Fe], [$\alpha_\text{ex}$/Fe], and [Al/Fe], most of the [X/H], [C/Mg], [Cr/Mg], and [Nd/Mg]. Of these, the [X/H] and [X/Mg] U-shaped trends are inverted compared to the [X/Fe] trends showing anti correlation. In addition, a similar trend is seen, albeit earlier, in the [Ti/Fe] and [Ti/Mg] gradients. With so many gradients sharing this feature, this is most likely due to a major event that shaped the overall chemical evolution of the LMC.

A majority of the age-[X/Fe] trends show an increase in the abundance ratio values that starts around the same time as the beginning of the U-shaped trend steepening, regardless of whether there is a U-shaped trend in their respective abundance gradient evolution. Also, this ramping up of abundance happens throughout the galaxy with the strongest effect in the central part of the LMC suggesting more relative star formation in the center. 
Some of the age-[X/Fe] trends have a turnover point $\sim$1$-$2 Gyr ago, which lines up with the extrema in the gradient evolutionary trends. This coincidence provides additional evidence of a major chemistry altering event. Due to the temporal proximity of the extreme point in the U-shaped gradient evolution trends and the increase in abundance seen in the age-[X/Fe] trends, these probably have a common cause. 

The LMC is known to have experienced a starburst $\sim$2 Gyr ago due to a close interaction between the LMC and SMC \citep[e.g.,][]{harris2009lmcsfh,nidever20lazy}. The LMC and SMC both show an increase in [$\alpha$/Fe] with respect to [Fe/H] for recent times, which indicates an increase in star formation in both galaxies. Based on the timing, this starburst event is the most likely cause for what is seen in the LMC abundance gradients and age-[X/Fe] trends. Galaxy interactions are known to induce bursts of star formation that have profound effects on the chemistry of the constituent galaxies. Observationally this has been seen in systems such as the interacting dwarf pair dm1647+12, which show increased star formation due to intergalactic interactions \citep{privon2017widespread}. In this particular pair the star formation is much clumpier and less centrally located, which the authors attribute to the low mass of the system. This could account for the fact that the LMC appears to have more centrally concentrated star formation versus dm1647+12. Further a steepening of a gradient after a galactic interaction suggests that metal poor gas may have been deposited onto the galaxy and/or continued enrichment in the center of galaxies \citep[e.g.,][]{sparre2022gas,buck2023impact}.




The MCs are not in isolation but are two satellites of the MW. It turns out that the infall or crossing time of the MCs was $\sim$1.5 -- 2.0 Gyr ago when they entered the MW potential \citep{Besla2007,Kallivayalil2013,gomez2015infall,patel2017orbits}. It is definitely possible that the interaction with the MW halo has shaped the evolution of the abundance gradients in the LMC. In \cite{Besla2012} it was shown that many structures with recent star formation in the LMC are mainly due to a recent interaction with the SMC and less so the MW. This is taken as evidence that the LMC-SMC interaction is indeed more important for the evolution of the abundance gradients in the LMC rather than any interaction with the MW, but does not mean that the MW had no affect. Any interaction so far would just have a smaller extent on the inner workings of the LMC. This will obviously change as the MCs merge with the MW.

As stated previously there is a possibility that radial migration has had an effect on the abundance gradients of the LMC. 
It is important to state that the evolution described herein is due to a conflation of both chemical and dynamical evolution. It is expected that older stars will be affected more by radial migration as they have been around longer, so the gradients calculated for these stars are less likely to match their birth gradients. Under the assumption that the strength of radial migration has remained mostly constant, stars with steeper birth abundance gradients will still have a relatively steep gradient compared to the other flattened gradients even though the value has changed. 

In \cite{lu2022migration} the affect of radial migration on the metallicity gradient for the MW was explored. In that work it was found that over time the metallicity gradient will flatten out, starting with the oldest stars. Interestingly it was found that the MW metallicity gradient steepened 11--8 Gyr ago and that corresponds to when it is thought that a former dwarf galaxy merged with the MW to  create the present Gaia-Sausage-Enceladus (GSE) structure. The steepening after the interaction/merger happens because metal poor gas is deposited into the MW ISM. Even with radial migration, a merger is still evident in the age-dependence of the metallicity gradient. The steepening in the MW is reminiscent of the LMC gradient steepening starting $\sim$2 Gyr ago,  implying that the SMC may have dumped metal poor gas onto the LMC. 

Also \cite{ratcliffe2023mwgrads} finds a similar behaviour for [Fe/H] where a steepening occurs when the GSE progenitor merged with the MW. That work also finds that interactions with the Sagittarius dwarf spheroidal also steepened the gradient when the general trend would have predicted a flattening. Finally in that work [X/H] gradients show this effect more strongly compared to [X/Fe] gradients of various elements. This is similar to herein with the gradients for [X/H] and [X/Fe].

In simulations of dwarf galaxies it has been shown that radial migration only plays a minor role in shaping metallicity gradients compared to larger galaxies \citep{schroyen2013simulations}. If this is the case, it can be inferred that the gradients calculated for the LMC are closer to their birth value than for large galaxies such as the MW or Andromeda, but this does not mean radial migration has not occurred.

\section{Summary}
\label{sec:lmc_summary}


This paper presents the radial abundance trends for 25 abundance ratios and the evolution of their respective gradients. The main results of this paper are as follows:

\begin{itemize}
    \item The LMC was found to have an overall metallicity gradient of $-$0.0380 $\pm$ 0.0022 dex/kpc, which agrees with several literature values \citep[e.g.,][]{cioni2009grad,feast2010lmcgrad,haschke2012metallicity}.
    \item Many of the gradients have a U-shaped trend (inverted or not) and share an extreme point  2.2--3.7 Gyr ago that corresponds to an interaction between the LMC and SMC \citep{nidever20lazy,hasselquist2021satellites}.
    \item Many age-[X/Fe] trends show an increase in abundance just before the known LMC-SMC interaction and some of these trends show a hump with a maximum at the time of the interaction, further pinpointing the peak in star formation from the burst to $\sim$2 Gyr ago.
\end{itemize}

Work like this can help guide and constrain our current galactic interaction simulations to better reflect what is seen in Magellanic-like systems. In addition gradients and age-[X/Fe] trends are useful tools for studying the chemical evolution of galaxies.




This paper is the second in a series of three. In Paper I of the series, ages are determined for RGB stars in the LMC and used in this work to determine the evolution of radial abundance trends. Paper III will analyze the same trends as this work but in the SMC. To obtain the evolution of these trends for the SMC,the same age-determination method in Paper I will be used.


\section*{Acknowledgements}


J.T.P. acknowledges support for this research from the National Science Foundation (AST-1908331) and the Montana Space Grant Consortium Graduate Fellowship.

D.L.N. also acknowledges supportfrom National Science Foundation (NSF) grant AST-1908331, while SRM and AA acknowledge NSF grant AST-1909497.
%
%

D.G. gratefully acknowledges the support provided by Fondecyt regular n. 1220264.
D.G. also acknowledges financial support from the Direcci\'{o}n de Investigaci\'{o}n y Desarrollo de
la Universidad de La Serena through the Programa de Incentivo a la Investigaci\'{o}n de
Acad\'{e}micos (PIA-DIDULS).

R.R.M. gratefully acknowledges support by the ANID BASAL project FB210003.

Funding for the Sloan Digital Sky Survey IV has been provided by the Alfred P. Sloan Foundation, the U.S. Department of Energy Office of Science, and the Participating Institutions. 

SDSS-IV acknowledges support and resources from the Center for High Performance Computing  at the University of Utah. The SDSS website is www.sdss.org.

SDSS-IV is managed by the Astrophysical Research Consortium for the Participating Institutions of the SDSS Collaboration including the Brazilian Participation Group, the Carnegie Institution for Science, Carnegie Mellon University, Center for Astrophysics | Harvard \& Smithsonian, the Chilean Participation Group, the French Participation Group, Instituto de Astrof\'isica de Canarias, The Johns Hopkins University, Kavli Institute for the Physics and Mathematics of the Universe (IPMU) / University of Tokyo, the Korean Participation Group, Lawrence Berkeley National Laboratory, Leibniz Institut f\"ur Astrophysik Potsdam (AIP), Max-Planck-Institut f\"ur Astronomie (MPIA Heidelberg), Max-Planck-Institut f\"ur Astrophysik (MPA Garching), Max-Planck-Institut f\"ur Extraterrestrische Physik (MPE), National Astronomical Observatories of China, New Mexico State University, New York University, University of Notre Dame, Observat\'ario Nacional / MCTI, The Ohio State University, Pennsylvania State University, Shanghai Astronomical Observatory, United Kingdom Participation Group, Universidad Nacional Aut\'onoma de M\'exico, University of Arizona, University of Colorado Boulder, University of Oxford, University of Portsmouth, University of Utah, University of Virginia, University of Washington, University of Wisconsin, Vanderbilt University, and Yale University.

This work has made use of data from the European Space Agency (ESA) mission
{\it Gaia} (\url{https://www.cosmos.esa.int/gaia}), processed by the {\it Gaia}
Data Processing and Analysis Consortium (DPAC,
\url{https://www.cosmos.esa.int/web/gaia/dpac/consortium}). Funding for the DPAC
has been provided by national institutions, in particular the institutions
participating in the {\it Gaia} Multilateral Agreement.

\textit{Software:} Astropy \citep{pricewhelan2018astropy, robitaille2013astropy}, Matplotlib \citep{hunter2007matplotlib}, NumPy \citep{harris2020numpy}, SciPy \citep{virtanen2020scipy}

\section*{Data Availability}

All APOGEE DR17 data used in this study is publicly available and can be found at: \url{https://www.sdss4.org/dr17/data_access/}.




\bibliographystyle{mnras}
\bibliography{bibliography} 




\appendix

\section{\texorpdfstring{$\nabla_R$}\,[X/H] and \texorpdfstring{$\nabla_R$}\,[X/M\lowercase{g}] Tables}
\label{app:lmc_grads_h_mg}


Here are the table for the [X/H] and [X/Mg] gradients, which do not appear in the main body of the paper.

\begin{table*}
	\centering
	\caption{A table of the gradients for each of the age bins using H as the fiducial element. Horizontal lines have been added marking the division of the previously defined groups of elements. In general descending down the table corresponds to an increase in atomic number. Each age bin has its own separate column.}
	\label{tab:lmc_evolve_grad_h}
	\begin{tabular}{cccccc}
        \hline
         & t $\leq$ 2.23 & 2.23 $<$ t $\leq$ 3.66 & 3.66 $<$ t $\leq$ 5.58 & 5.58 $<$ t $\leq$ 8.36 & 8.36 $<$ t \\
        Element & $\nabla_\text{R}$ & $\nabla_\text{R}$ & $\nabla_\text{R}$ & $\nabla_\text{R}$ & $\nabla_\text{R}$ \\
         & (dex/kpc) & (dex/kpc) & (dex/kpc) & (dex/kpc) & (dex/kpc) \\
        \hline 
        {[C/H]} & $-$0.0893 $\pm$ 0.0027 & $-$0.0113 $\pm$ 0.0042 & $-$0.0359 $\pm$ 0.0047 & $-$0.0510 $\pm$ 0.0056 & $-$0.0680 $\pm$ 0.0077 \\
        {[N/H]} & $-$0.0803 $\pm$ 0.0054 & $-$0.0080 $\pm$ 0.0029 & $-$0.0255 $\pm$ 0.0031 & $-$0.017 $\pm$ 0.0028 & $-$0.0222 $\pm$ 0.0047 \\
        {[(C+N)/H]} & $-$0.0732 $\pm$ 0.0038 & $-$0.0694 $\pm$ 0.0013 & $-$0.0325 $\pm$ 0.0040 & $-$0.0405 $\pm$ 0.0049 & $-$0.0520 $\pm$ 0.0063 \\
        \hline
        {[O/H]} & $-$0.0439 $\pm$ 0.0021 & $-$0.0491 $\pm$ 0.0011 & $-$0.0181 $\pm$ 0.0034 & $-$0.0216 $\pm$ 0.0041 & $-$0.0323 $\pm$ 0.0046 \\
        {[Mg/H]} & $-$0.0309 $\pm$ 0.0024 & $-$0.0135 $\pm$ 0.0028 & $-$0.0229 $\pm$ 0.0033 & $-$0.0297 $\pm$ 0.0042 & $-$0.0308 $\pm$ 0.0049 \\
        {[Si/H]} & $-$0.0432 $\pm$ 0.0025 & $-$0.0248 $\pm$ 0.0019 & $-$0.0125 $\pm$ 0.0032 & $-$0.0194 $\pm$ 0.0039 & $-$0.0259 $\pm$ 0.0048 \\
        {[S/H]} & $-$0.0157 $\pm$ 0.0036 & $-$0.0064 $\pm$ 0.0058 & $-$0.0129 $\pm$ 0.0058 & $-$0.0125 $\pm$ 0.0071 & $-$0.0326 $\pm$ 0.0116 \\
        {[Ca/H]} & $-$0.0362 $\pm$ 0.0027 & $-$0.0131 $\pm$ 0.0023 & $-$0.0181 $\pm$ 0.003 & $-$0.0266 $\pm$ 0.004 & $-$0.0334 $\pm$ 0.0051 \\
        {[Ti/H]} & $-$0.0294 $\pm$ 0.0034 & $-$0.0385 $\pm$ 0.0045 & $-$0.0522 $\pm$ 0.0121 & 0.0603 $\pm$ 0.0086 & 0.0476 $\pm$ 0.0151 \\
        {[$\alpha$/H]} & $-$0.0451 $\pm$ 0.0022 & $-$0.0283 $\pm$ 0.0022 & $-$0.0201 $\pm$ 0.0033 & $-$0.025 $\pm$ 0.0041 & $-$0.0348 $\pm$ 0.0052 \\
        {[$\alpha_\text{h}$/H]} & $-$0.0479 $\pm$ 0.002 & $-$0.0154 $\pm$ 0.0027 & $-$0.0210 $\pm$ 0.0033 & $-$0.0283 $\pm$ 0.0042 & $-$0.0315 $\pm$ 0.0047 \\
        {[$\alpha_\text{ex}$/H]} & $-$0.0451 $\pm$ 0.0026 & $-$0.0531 $\pm$ 0.0014 & $-$0.0205 $\pm$ 0.0100 & 0.0491 $\pm$ 0.0082 & 0.0424 $\pm$ 0.0130 \\
        \hline
        {[Na/H]} & $-$0.0607 $\pm$ 0.0038 & $-$0.0439 $\pm$ 0.0043 & $-$0.0100 $\pm$ 0.0056 & $-$0.0301 $\pm$ 0.0015 & $-$0.0071 $\pm$ 0.0058 \\
        {[Al/H]} & $-$0.0468 $\pm$ 0.0038 & $-$0.0595 $\pm$ 0.0015 & $-$0.0149 $\pm$ 0.0048 & $-$0.0122 $\pm$ 0.0068 & $-$0.0101 $\pm$ 0.0083 \\
        {[K/H]} & $-$0.0264 $\pm$ 0.0030 & $-$0.0108 $\pm$ 0.0035 & $-$0.0185 $\pm$ 0.0040 & $-$0.0276 $\pm$ 0.0044 & $-$0.0265 $\pm$ 0.0060 \\
        \hline
        {[V/H]} & $-$0.0190 $\pm$ 0.0038 & $-$0.0143 $\pm$ 0.0037 & $-$0.0362 $\pm$ 0.0037 & $-$0.0529 $\pm$ 0.0057 & $-$0.0647 $\pm$ 0.0086 \\
        {[Cr/H]} & $-$0.0378 $\pm$ 0.0035 & 0.0005 $\pm$ 0.0035 & $-$0.086 $\pm$ 0.0017 & $-$0.0267 $\pm$ 0.0073 & $-$0.0418 $\pm$ 0.0087 \\
        {[Mn/H]} & $-$0.0627 $\pm$ 0.0038 & $-$0.0049 $\pm$ 0.0053 & $-$0.0267 $\pm$ 0.0112 & 0.0279 $\pm$ 0.014 & 0.0644 $\pm$ 0.0136 \\
        {[Fe/H]} & $-$0.0354 $\pm$ 0.0031 & $-$0.0149 $\pm$ 0.002 & $-$0.0179 $\pm$ 0.0031 & $-$0.0286 $\pm$ 0.0042 & $-$0.0439 $\pm$ 0.0049 \\
        {[Co/H]} & $-$0.0394 $\pm$ 0.003 & $-$0.0443 $\pm$ 0.0027 & $-$0.0241 $\pm$ 0.0041 & $-$0.0333 $\pm$ 0.0052 & $-$0.0499 $\pm$ 0.0066 \\
        {[Ni/H]} & $-$0.0311 $\pm$ 0.0026 & $-$0.0085 $\pm$ 0.0026 & $-$0.0274 $\pm$ 0.0027 & $-$0.0304 $\pm$ 0.0042 & $-$0.0381 $\pm$ 0.0055 \\
        \hline
        {[Ce/H]} & $-$0.0875 $\pm$ 0.0180 & $-$0.0142 $\pm$ 0.0090 & $-$0.0232 $\pm$ 0.0141 & $-$0.0096 $\pm$ 0.0077 & $-$0.0270 $\pm$ 0.0160 \\
        {[Nd/H]} & $-$0.0367 $\pm$ 0.0109 & $-$0.0163 $\pm$ 0.0091 & $-$0.0061 $\pm$ 0.0079 & $-$0.0205 $\pm$ 0.0086 & $-$0.0114 $\pm$ 0.0104 \\
    \end{tabular}
\end{table*}

\begin{table*}
	\centering
	\caption{A table of the gradients for each of the age bins using H as the fiducial element. Horizontal lines have been added marking the division of the previously defined groups of elements. In general descending down the table corresponds to an increase in atomic number. Each age bin has its own separate column.}
	\label{tab:lmc_evolve_grad_mg}
	\begin{tabular}{cccccc}
        \hline
         & t $\leq$ 2.23 & 2.23 $<$ t $\leq$ 3.66 & 3.66 $<$ t $\leq$ 5.58 & 5.58 $<$ t $\leq$ 8.36 & 8.36 $<$ t \\
        Element & $\nabla_\text{R}$ & $\nabla_\text{R}$ & $\nabla_\text{R}$ & $\nabla_\text{R}$ & $\nabla_\text{R}$ \\
         & (dex/kpc) & (dex/kpc) & (dex/kpc) & (dex/kpc) & (dex/kpc) \\
        \hline 
        {[C/Mg]} & -0.0296 $\pm$ 0.0023 & -0.0073 $\pm$ 0.002 & -0.012 $\pm$ 0.0019 & -0.0223 $\pm$ 0.0025 & -0.0276 $\pm$ 0.0035 \\
        {[N/Mg]} & -0.0514 $\pm$ 0.0035 & 0.0019 $\pm$ 0.0013 & 0.0016 $\pm$ 0.0012 & 0.0044 $\pm$ 0.0016 & 0.0081 $\pm$ 0.002 \\
        {[(C+N)/Mg]} & -0.0368 $\pm$ 0.0024 & -0.0060 $\pm$ 0.0014 & -0.0087 $\pm$ 0.0013 & -0.0114 $\pm$ 0.0014 & -0.0091 $\pm$ 0.0019 \\
        \hline
        {[O/Mg]} & -0.0028 $\pm$ 0.0009 & 0.0041 $\pm$ 0.0007 & 0.0022 $\pm$ 0.0008 & 0.0042 $\pm$ 0.001 & 0.0035 $\pm$ 0.0013 \\
        {[Mg/H]} & $-$0.0309 $\pm$ 0.0024 & $-$0.0135 $\pm$ 0.0028 & $-$0.0229 $\pm$ 0.0033 & $-$0.0297 $\pm$ 0.0042 & $-$0.0308 $\pm$ 0.0049 \\
        {[Si/Mg]} & -0.0065 $\pm$ 0.0014 & 0.0087 $\pm$ 0.0013 & 0.0076 $\pm$ 0.0008 & 0.0089 $\pm$ 0.0011 & 0.0095 $\pm$ 0.0018 \\
        {[S/Mg]} & 0.0075 $\pm$ 0.0038 & 0.0159 $\pm$ 0.0056 & 0.0120 $\pm$ 0.0062 & 0.0383 $\pm$ 0.0047 & -0.0045 $\pm$ 0.0112 \\
        {[Ca/Mg]} & -0.0085 $\pm$ 0.0013 & 0.0000 $\pm$ 0.0012 & 0.0013 $\pm$ 0.0012 & 0.0013 $\pm$ 0.0014 & -0.0022 $\pm$ 0.0016 \\
        {[Ti/Mg]} & -0.0037 $\pm$ 0.0019 & -0.007 $\pm$ 0.0029 & -0.0229 $\pm$ 0.0026 & -0.0023 $\pm$ 0.0054 & 0.0170 $\pm$ 0.0044 \\
        {[$\alpha$/Mg]} & -0.0054 $\pm$ 0.0009 & 0.0015 $\pm$ 0.0009 & -0.0005 $\pm$ 0.0008 & 0.0007 $\pm$ 0.0011 & 0.0004 $\pm$ 0.0012 \\
        {[$\alpha_\text{h}$/Mg]} & -0.0014 $\pm$ 0.0004 & 0.0021 $\pm$ 0.0004 & 0.0011 $\pm$ 0.0004 & 0.0021 $\pm$ 0.0005 & 0.0017 $\pm$ 0.0006 \\
        {[$\alpha_\text{ex}$/Mg]} & -0.0068 $\pm$ 0.0012 & -0.0002 $\pm$ 0.0018 & -0.0053 $\pm$ 0.0017 & 0.0016 $\pm$ 0.0020 & 0.0056 $\pm$ 0.0026 \\
        \hline
        {[Na/Mg]} & -0.0173 $\pm$ 0.0033 & 0.0020 $\pm$ 0.0041 & 0.0141 $\pm$ 0.0038 & 0.0079 $\pm$ 0.0042 & 0.0131 $\pm$ 0.0046 \\
        {[Al/Mg]} & -0.0172 $\pm$ 0.0021 & 0.0044 $\pm$ 0.0017 & 0.0019 $\pm$ 0.0018 & 0.0022 $\pm$ 0.0015 & 0.0024 $\pm$ 0.0021 \\
        {[K/Mg]} & 0.0029 $\pm$ 0.0021 & 0.0020 $\pm$ 0.0016 & 0.0054 $\pm$ 0.0017 & 0.0007 $\pm$ 0.0018 & 0.0037 $\pm$ 0.0024 \\
        \hline
        {[V/Mg]} & -0.0032 $\pm$ 0.003 & -0.0076 $\pm$ 0.0020 & -0.0121 $\pm$ 0.002 & -0.0173 $\pm$ 0.0026 & -0.0212 $\pm$ 0.004 \\
        {[Cr/Mg]} & -0.0172 $\pm$ 0.0025 & -0.0006 $\pm$ 0.0028 & -0.0049 $\pm$ 0.0028 & -0.0066 $\pm$ 0.0036 & -0.0099 $\pm$ 0.0041 \\
        {[Mn/Mg]} & -0.0284 $\pm$ 0.0018 & -0.0035 $\pm$ 0.0020 & -0.013 $\pm$ 0.0012 & 0.0041 $\pm$ 0.004 & 0.0086 $\pm$ 0.0043 \\
        {[Fe/Mg]} & -0.0685 $\pm$ 0.0052 & -0.0388 $\pm$ 0.0048 & -0.0417 $\pm$ 0.0062 & -0.0575 $\pm$ 0.0084 & -0.0727 $\pm$ 0.0103 \\
        {[Co/Mg]} & -0.0060 $\pm$ 0.0017 & -0.0059 $\pm$ 0.0019 & -0.0050 $\pm$ 0.0018 & -0.0054 $\pm$ 0.0018 & -0.0154 $\pm$ 0.0027 \\
        {[Ni/Mg]} & -0.0072 $\pm$ 0.0012 & -0.0013 $\pm$ 0.0013 & -0.0019 $\pm$ 0.0013 & -0.0027 $\pm$ 0.0016 & -0.0048 $\pm$ 0.0017 \\
        \hline
        {[Ce/Mg]} & -0.0507 $\pm$ 0.0123 & 0.0103 $\pm$ 0.0088 & 0.0017 $\pm$ 0.0076 & 0.0075 $\pm$ 0.0081 & 0.0018 $\pm$ 0.0075 \\
        {[Nd/Mg]} & 0.0020 $\pm$ 0.0099 & -0.0141 $\pm$ 0.0082 & 0.0022 $\pm$ 0.0097 & -0.0001 $\pm$ 0.0087 & 0.0188 $\pm$ 0.0082 \\
    \end{tabular}
\end{table*}

\section{Radius-[X/F\lowercase{e}]-Age Trends}
\label{app:lmc_age_xfe_radius_trends}


Here are the radius-age trends for the $\alpha$-elements, odd-Z elements, iron peak elements and neutron capture elements, which do not appear in the main body of the paper.

\begin{figure*}
    \centering
    \includegraphics[width=0.949\textwidth]{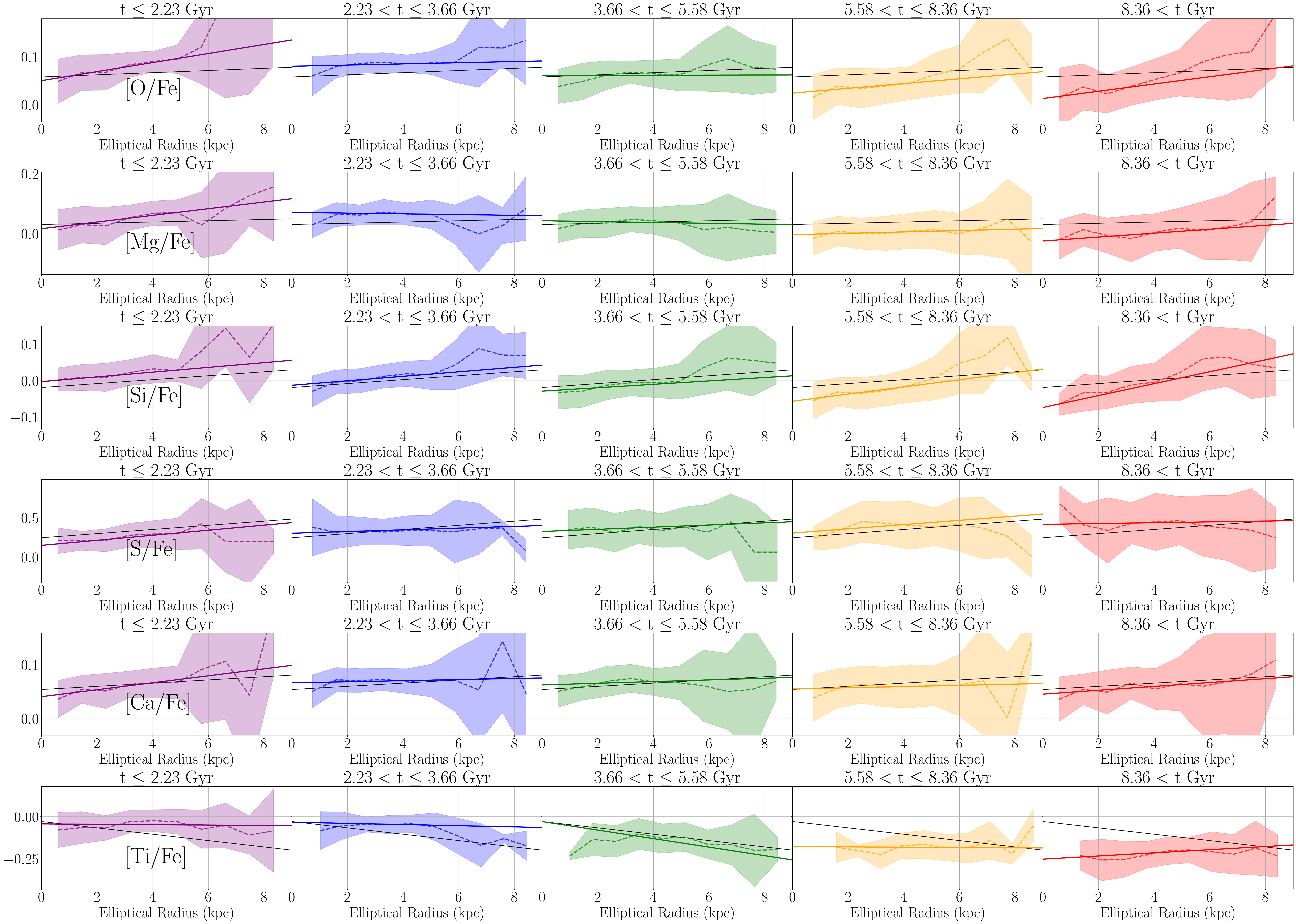}
    \caption{The same as for Figure~\ref{fig:lmc_cn_radxfe}, but with the individual $\alpha$-element abundances.}
    \label{fig:lmc_ind_alpha_radxfe}
\end{figure*}

\begin{figure*}
    \centering
    \includegraphics[width=0.95\textwidth]{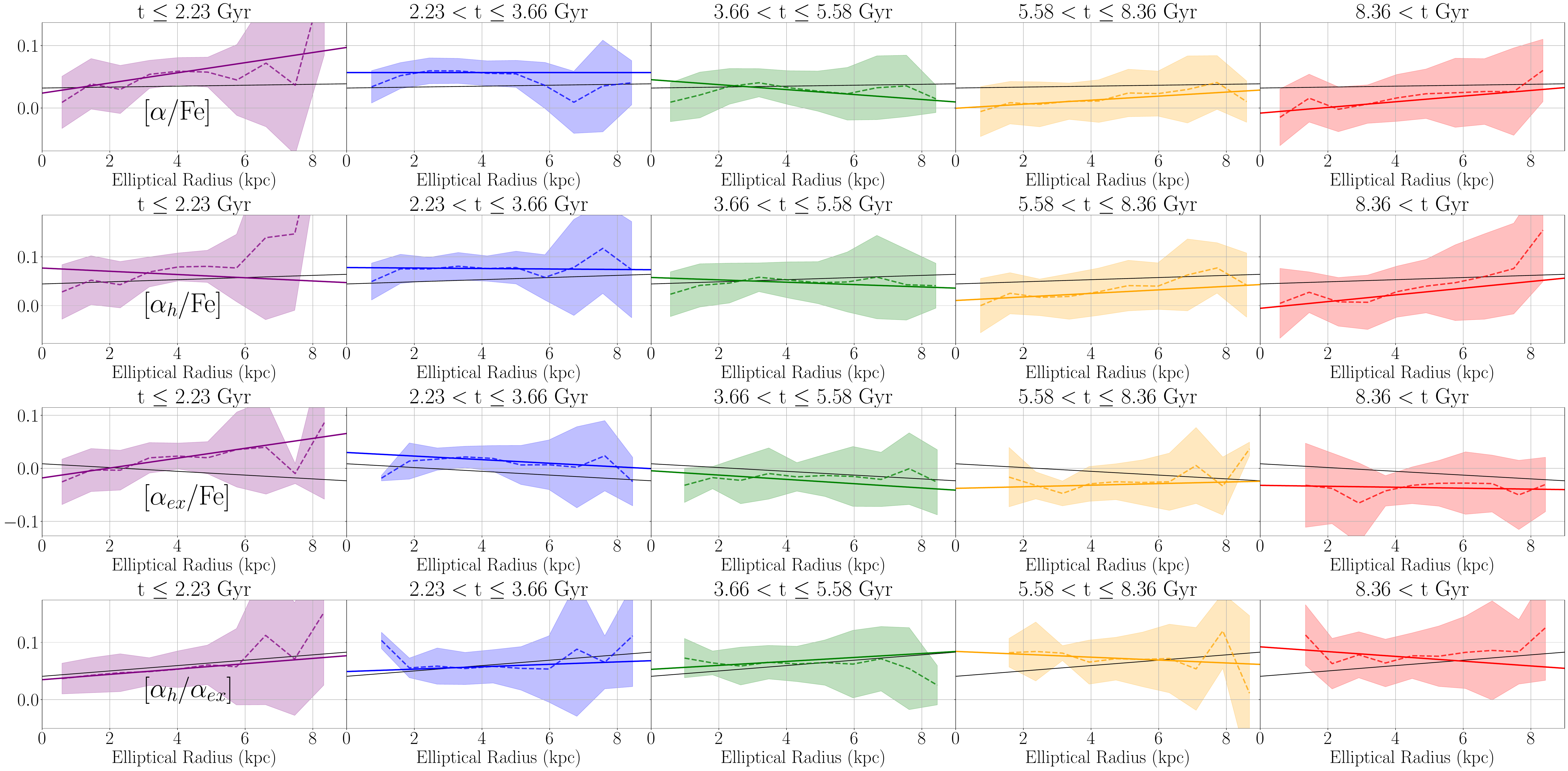}
    \caption{The radial abundance trends for the composite $\alpha$-element abundances.}
    \label{fig:lmc_comb_alpha_radxfe}
\end{figure*}

\begin{figure*}
    \centering
    \includegraphics[width=0.95\textwidth]{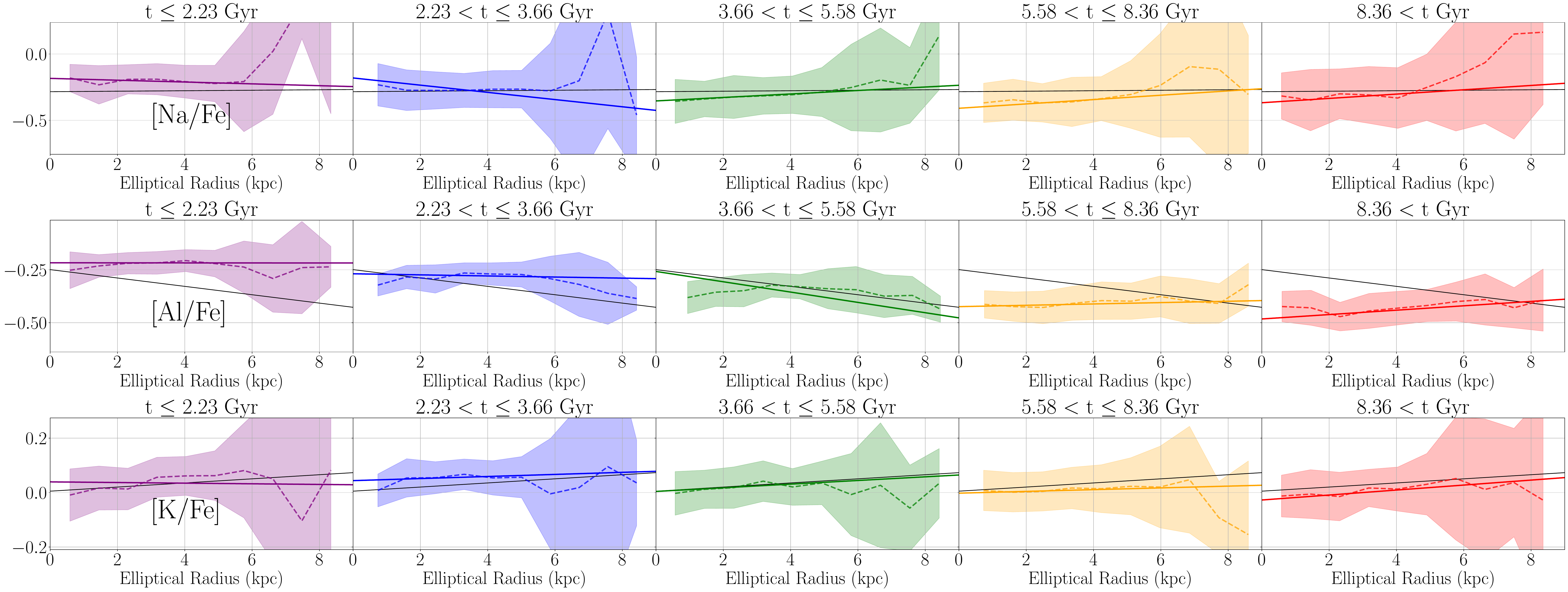}
    \caption{The radial abundance trends for the odd-Z abundances.}
    \label{fig:lmc_oddz_radxfe}
\end{figure*}

\begin{figure*}
    \centering
    \includegraphics[width=0.95\textwidth]{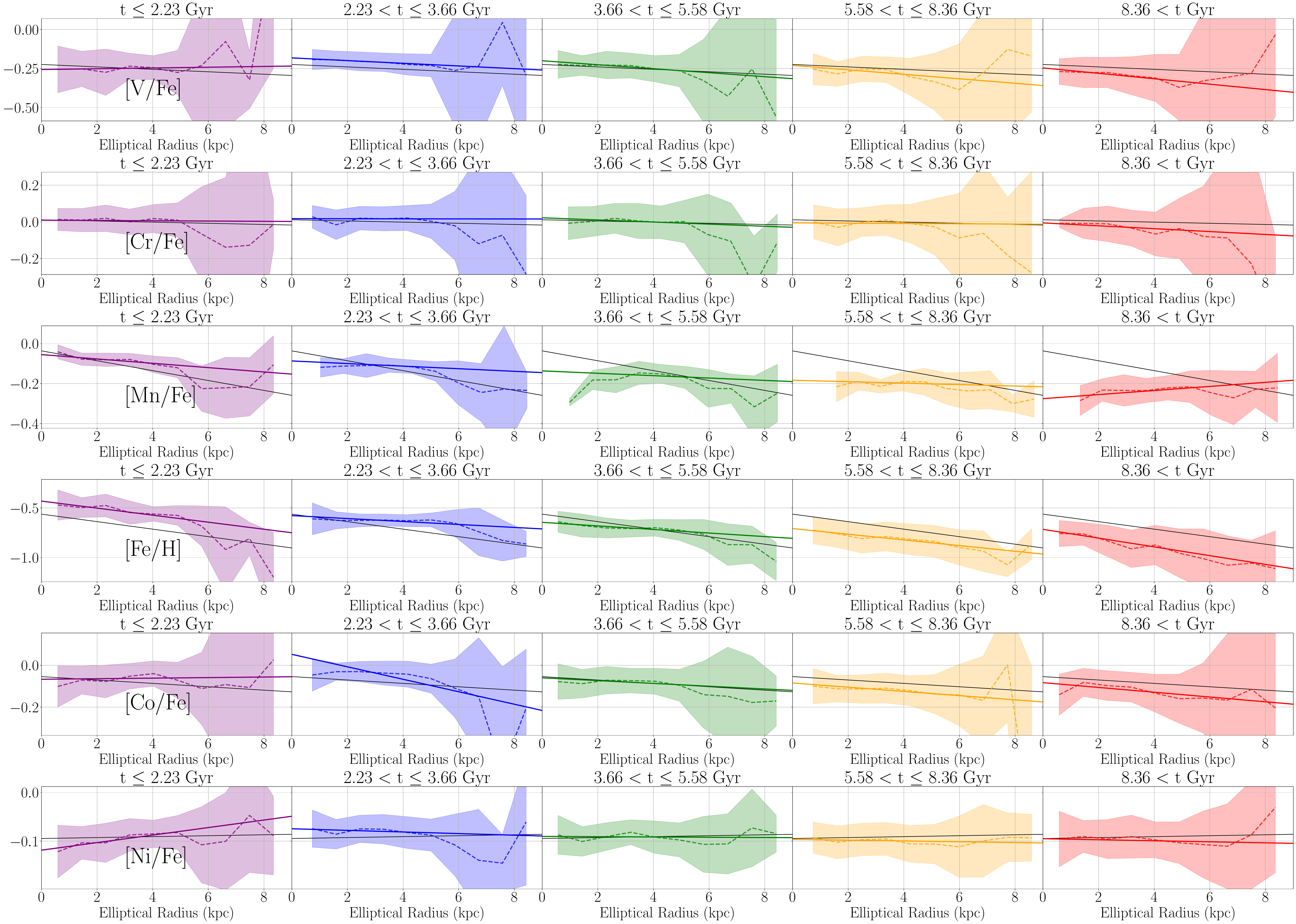}
    \caption{The radial abundance trends for the iron peak abundances.}
    \label{fig:lmc_ironpeak_radxfe}
\end{figure*}

\begin{figure*}
    \centering
    \includegraphics[width=0.95\textwidth]{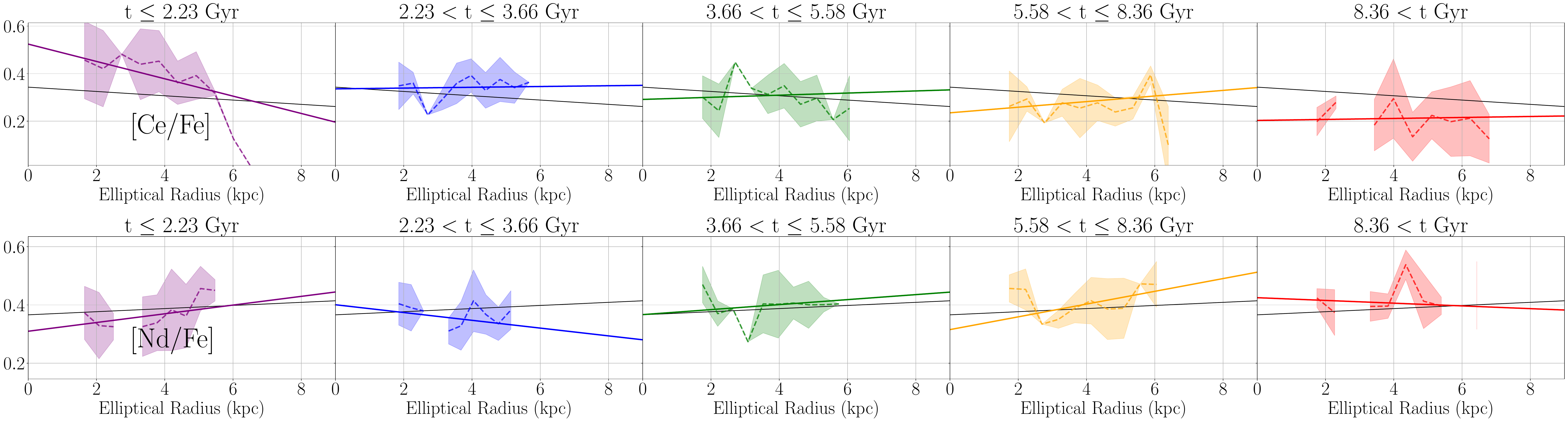}
    \caption{The radial abundance trends for the neutron capture abundances.}
    \label{fig:lmc_sr_radxfe}
\end{figure*}

\section{age-[X/F\lowercase{e}] Trends}
\label{app:lmc_age_xfe_trends}


Here are the age-[X/Fe] trends for the $\alpha$-elements, odd-Z elements, iron peak elements and neutron capture elements, which do not appear in the main body of the paper.

\begin{figure*}
    \centering
    \includegraphics[width=0.85\textwidth,height=0.9\textheight]{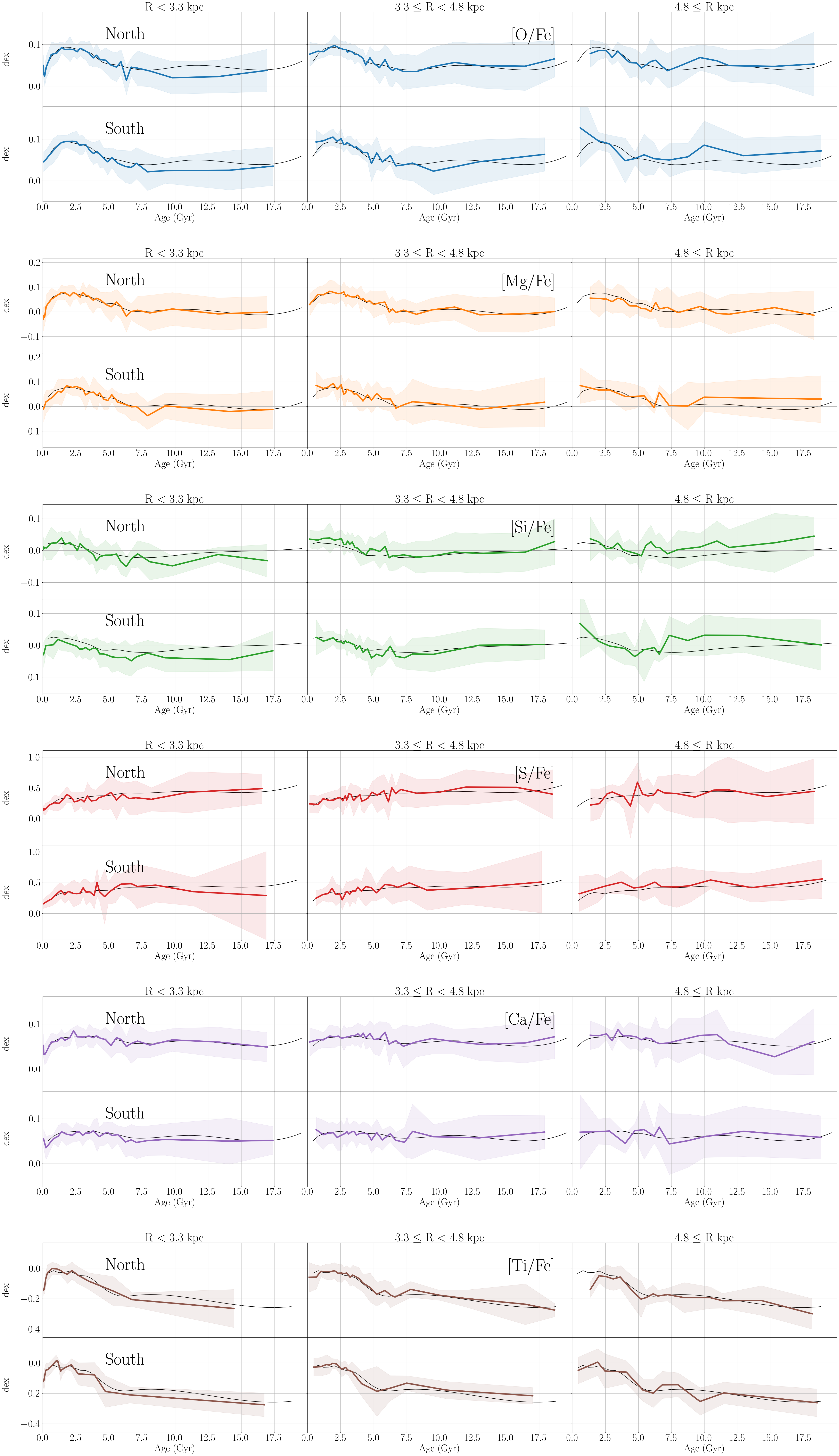}
    \caption{The age-[X/Fe] trends for the individual $\alpha$-elements.}
    \label{fig:lmc_ind_alpha_axfe}
\end{figure*}

\begin{figure*}
    \centering
    \includegraphics[width=0.875\textwidth]{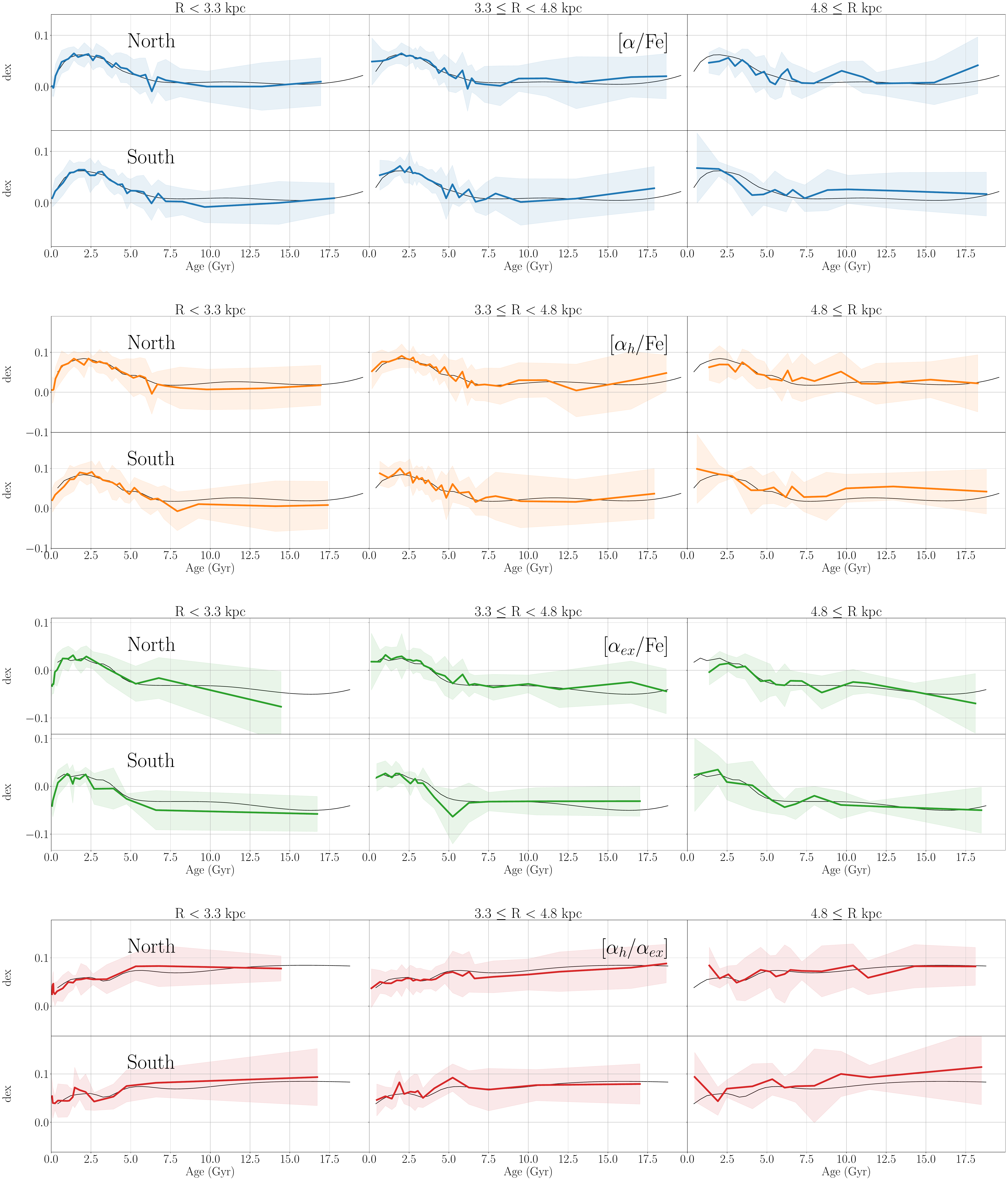}
    \caption{The age-[X/Fe] trends for the composite $\alpha$-elements.}
    \label{fig:lmc_comb_alpha_axfe}
\end{figure*}

\begin{figure*}
    \centering
    \includegraphics[width=0.95\textwidth]{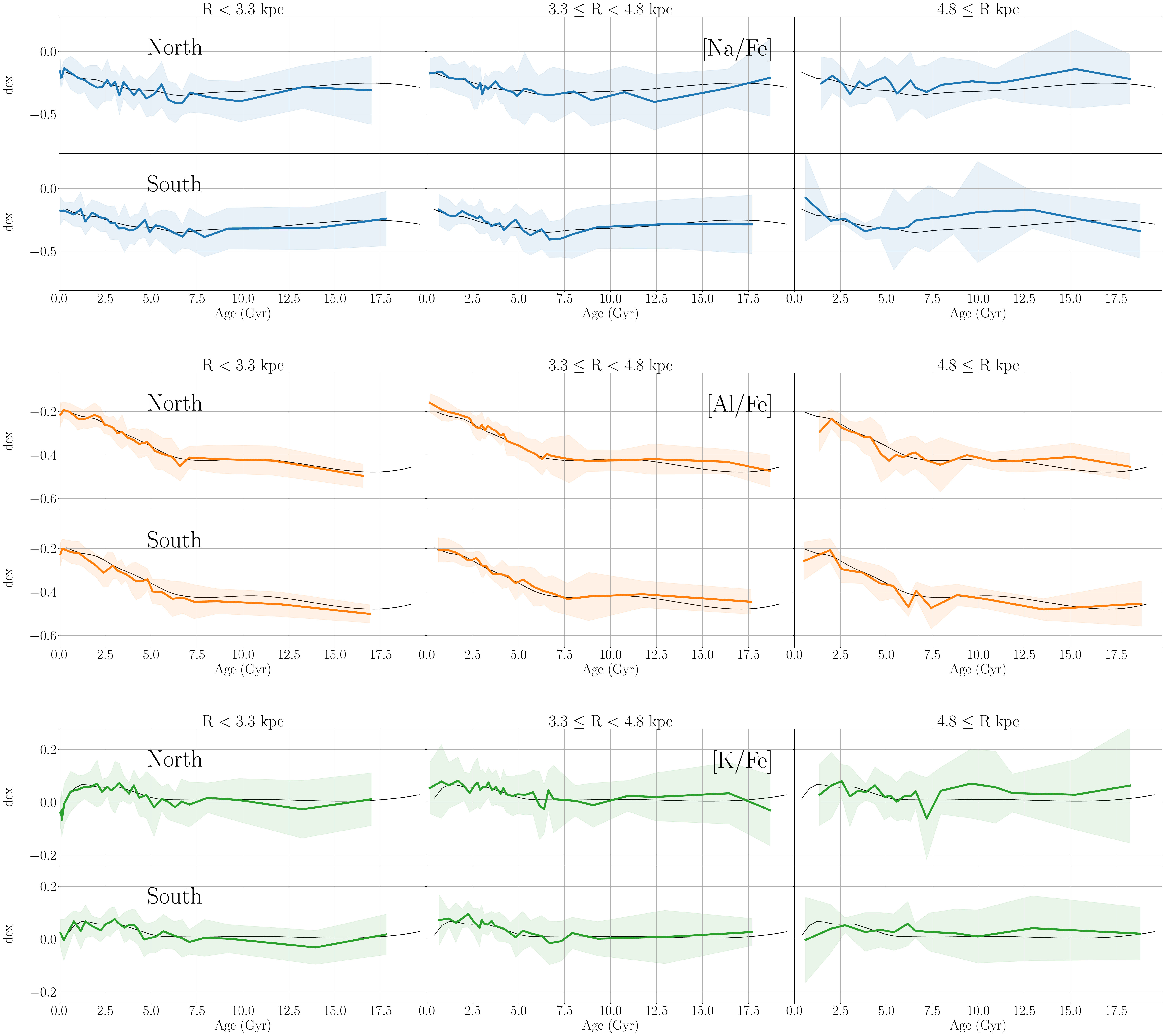}
    \caption{The age-[X/Fe] trends for the odd-Z elements.}
    \label{fig:lmc_oddz_axfe}
\end{figure*}

\begin{figure*}
    \centering
    \includegraphics[width=0.9\textwidth,height=0.9\textheight]{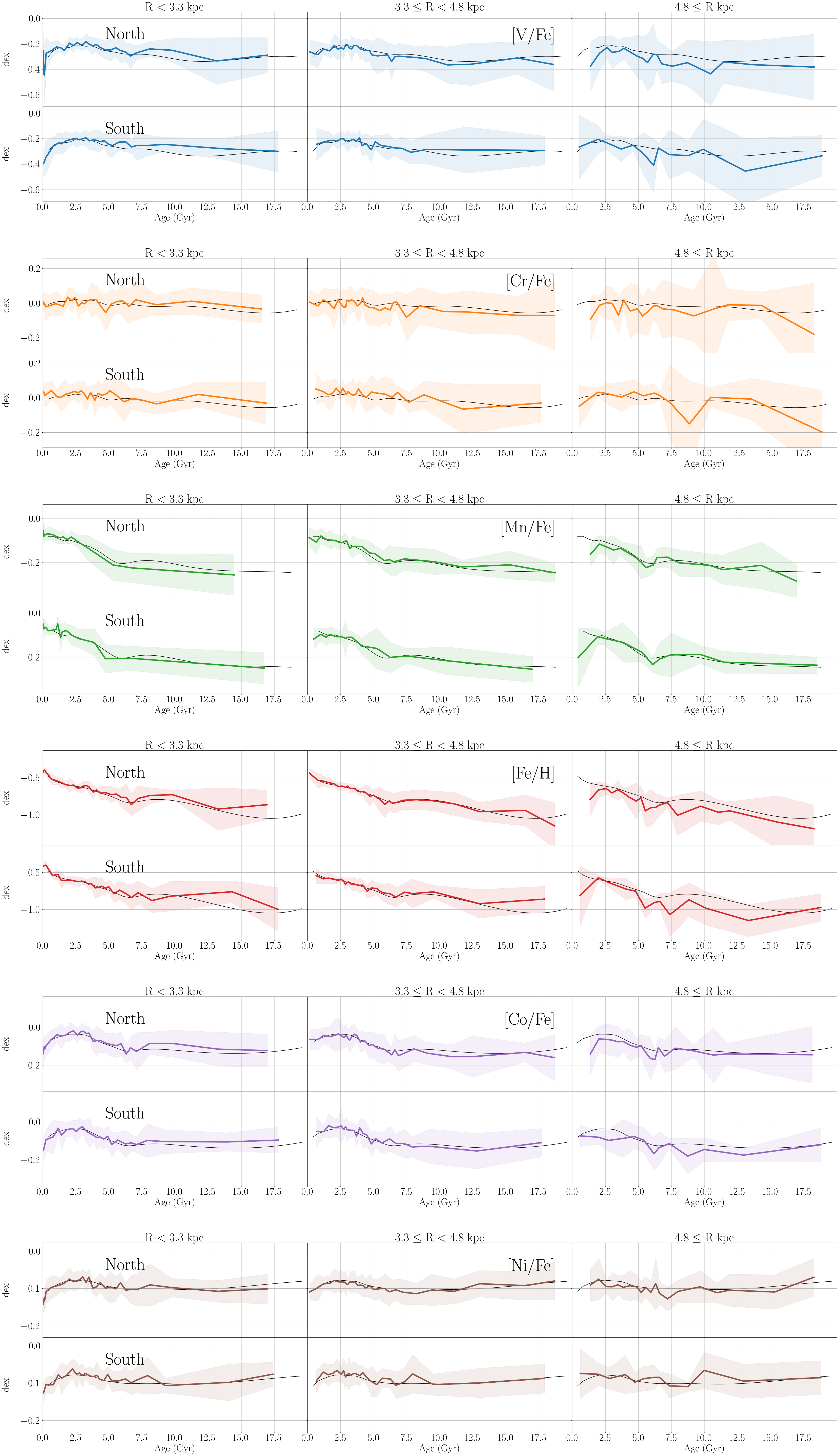}
    \caption{The age-[X/Fe] trends for the iron peak elements.}
    \label{fig:lmc_ironpeak_axfe}
\end{figure*}

\begin{figure*}
    \centering
    \includegraphics[width=0.95\textwidth]{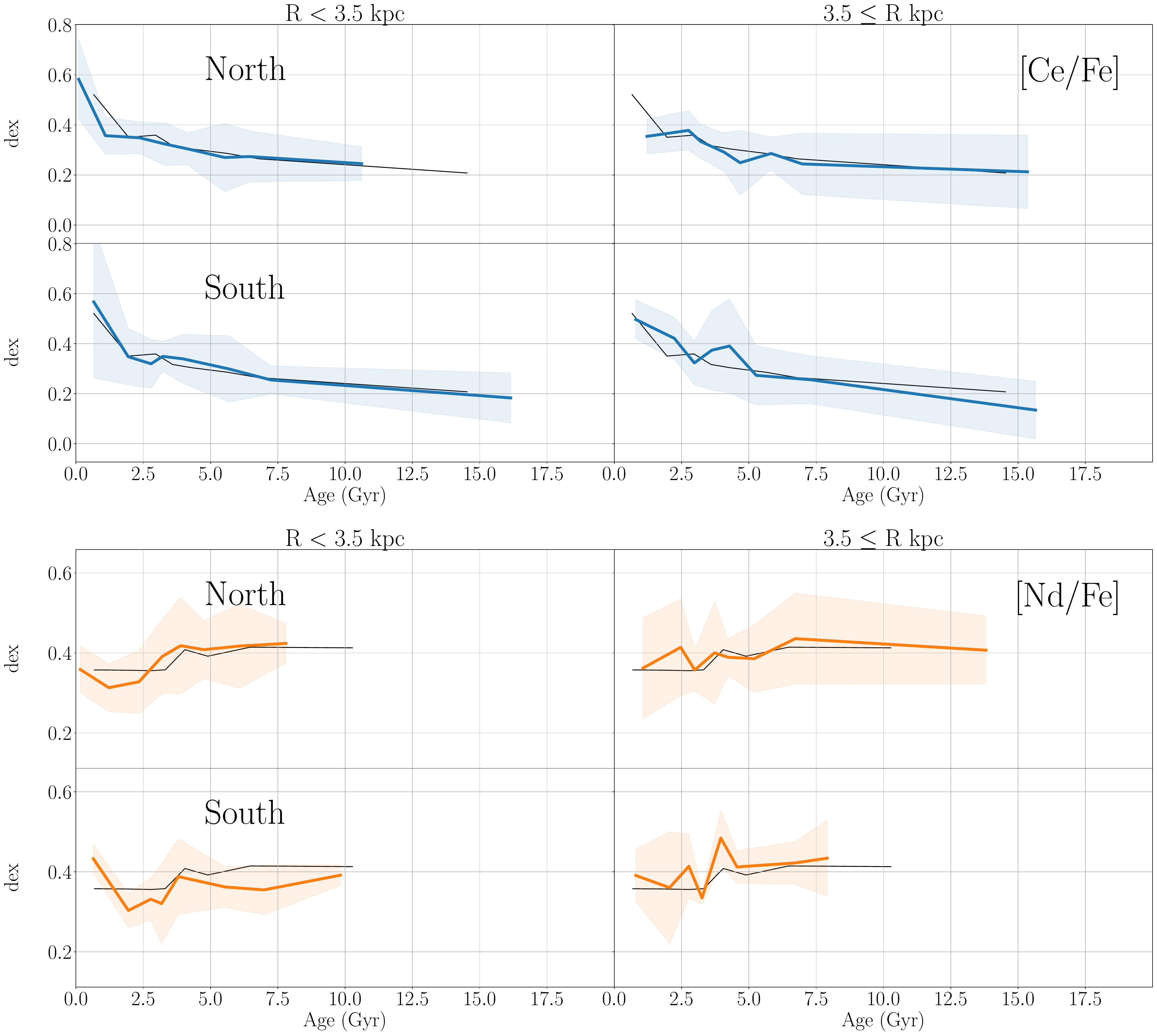}
    \caption{The age-[X/Fe] trends for the neutron capture elements.}
    \label{fig:lmc_sr_axfe}
\end{figure*}



\bsp	
\label{lastpage}
\end{document}